\documentclass [preprint] {revtex4}
\usepackage{graphicx}
\usepackage{amssymb}

\def\lp {\left( }
\def\rp {\right) }
\def\lb {\left[ }
\def\rb {\right] }
\def\lc {\left\{ }
\def\rc {\right\} }
\def\ra {\rangle }
\def\la {\langle }

\def\rar {\rightarrow}
\def\lrar {\leftrightarrow}

\def\beq{\begin{equation}}
\def\eeq{\end{equation}}
\def\bea{\begin{eqnarray}}
\def\eea{\end{eqnarray}}
\def\nn {\nonumber}
\def\ni {\noindent}

\def\cd {\!\cdot\!}

\def\Qt {\tilde{Q}}
\def\dr {\partial }
\def\Ob {\bar{\Omega}}
\def\Kb {\bar{K}}
\def\rtw {\sqrt{2}}
\def\rth {\sqrt{3}}
\def\sp {\!+\!}
\def\sm {\!-\!}

\def\mr2 {m_\rho^2 }

\def\cK {{\cal{K}}}
\def\cKb {\bar{{\cal{K}}}}
\def\cL {{\cal L}}

\def\a{\alpha}
\def\b{\beta}
\def\d{\delta}
\def\e{\epsilon}
\def\g{\gamma}
\def\D {\Delta}
\def\f {\phi}
\def\G {\Gamma}

\def\l{\lambda}
\def\m{\mu}
\def\n{\nu}
\def\o{\omega}

\def\p {\pi}
\def\r{\rho}
\def\s{\sigma}

\def\th {\theta}

\def\S{\Sigma}

\def\x {\chi}
\def\y {\eta}
\def\z {\zeta}
\def\w{\omega}

\def\dkkk{ D^+ \to K^+ K^- K^+}

\begin{document}

\title{Multi-Meson Model applied to $\dkkk$}

\author{R. T. Aoude$^a$, P. C. Magalh\~aes$^a$, A. C. dos Reis$^a$ and  M. R. Robilotta$^b$}
\affiliation{$^a$Centro Brasileiro de Pesquisas F\'isicas  --  CBPF, 
Rio de Janeiro, RJ, Brazil \\
$^b$ Instituto de F\'{\i}sica, Universidade de S\~{a}o Paulo,  S\~{a}o Paulo, SP, Brazil}

 \email[]{pmagalhaes@cbpf.br}
\date{\today }

\begin{abstract}

Matrix elements of weak currents involving light multi-meson states 
are important in many hadronic decays of both heavy leptons and heavy mesons.
In this paper we focus on the latter case where the current size of the data set  demands better models.
The specific case of three-kaon weak matrix elements 
is considered and
expressed as a relatively simple structure,
which generalizes naturally the concept of form factor. 
We propose a model for the decay  $\dkkk$ as an alternative to isobar model, with
free parameter predicted by the theory to be fine-tuned by  a fit to data.
An important qualitative outcome is  that 
we encompass naturally all final states topologies,
which involve necessarily proper multi-particle structures 
and cannot be decomposed into simpler two-body processes.  This aspect represents  a significant improvement when compared to isobar
model, often employed in analyses of heavy-meson decay data.  

\end{abstract}

\pacs{...}

\maketitle

\section{introduction}

Three-body, nonleptonic decays of heavy-flavoured mesons
are sequential processes, dominated by intermediate 
resonant states.  For this reason, these decays
have been extensively used to study light meson spectroscopy. 
For the same reason, these decays have been also used  for direct CP violation searches. The phase variation of
the resonances provides the strong phase difference required
for CP violation to occur. 

The determination of the resonant structure of three-body
decays requires a full amplitude analysis. The key issue 
in such analysis is the modelling of the decay amplitude.
The usual experimental approach is to represent the decay amplitude as
a coherent sum of resonant amplitudes, $\mathcal{A}=\sum c_k A_k$. 
The relative contribution of the different resonant amplitudes
are derived from the complex coefficients $c_k$, which
are the usual fit parameters. This is known as the isobar model.

The isobar model provides an effective description of the
Dalitz plot of many different final states,
for data sets  up to $\mathcal{O}(10^4)$ events \cite{asner}.
However, the analysis of the gigantic samples of $B$ and $D$ decays
collected by the LHC experiments demands better models.

A particularly important issue is the representation of
  the nonresonant component of the decay amplitude. 
 In general, the nonresonant amplitude used to fit data is parametrized by  ad hoc
  functions that are not compromised with any theory.
 This component is typically small in $D$ meson decays, and it is usually
assumed to be constant across the Dalitz plot \cite{asner}. In $B$ decays, 
however, where the contribution of the nonresonant component may be large, 
empirical formulas have been 
used to fit the data. These formulas may provide an effective 
representation of the available data, but the lack of theoretical
justification and the interplay between the nonresonant amplitude 
and the broad scalar resonances at low  masses make  the 
interpretation of the results rather difficult.
Irrespective to the size of the nonresonant contribution,
a proper formulation for this amplitude is in order.

In this paper, a new approach for the decay amplitude is
presented, as an alternative to the isobar model. 
The formalism is applicable to decays of charged mesons --
$D^+$, $D^+_s$ and $B^+$ -- into three kaons. These decays
may proceed through a common topology: the annihilation diagram,
in the language of the quark diagrammatic approach proposed by 
Chau\cite{Chau}.

In  this specific topology, the decay amplitude
may be written as 
\[ 
\mathcal{A} =
\langle (KKK)^+|A_{\mu}|0\rangle \langle 0|A_{\mu}|M^+\rangle
\]
where $A_{\mu}$ is the axial weak current.
The second term in the right-hand side corresponds to
the weak vertex and depends on the quark content of the initial 
meson $M^+$. The firt term of the decay amplitude,
$\langle (KKK)^+|A_{\mu}|0\rangle$, hereafter referred 
to as the Multi-Meson Model, or Triple-M, has an universal character,
and is the main subject of this work.

This paper is focused, without any loss of generality, on the 
$D^+ \to K^-K^+K^+$ decay, for which the annihilation diagram is 
expected to be the dominant  mechanism.
Other possible topologies for this decay involve rescattering, and will be the subject  of  a future publication. 
The formalism can be easily extended to the decays of the $D^+_s$ and 
$B^+$ mesons, together with  other topologies that must be considered.

The Triple-M amplitude contains three components: $(KKK)^+$ nonresonant,
the $f_0(980) K^+$ and $\phi K^+$, derived from a chiral effective 
theory and dressed with coupled channels where appropriate.

In the Triple-M amplitude,
the relative contribution and phase of each component is fixed
by theory, and this represent an important difference with the isobar model. 
 There are only three parameters in the Triple-M related to the
 mass of $f_0$ and its couplings to light pseudo-scalars. At present, the values of
these parameters are estimated by  theory but, ultimately, they should be determined by fits to the data.

The model, in its present version, does not include
 three-body final state interactions 
(FSI). These FSI have been shown to play a significant role in
the $D^+ \to K^-\pi^+\pi^+$ decay \cite{BR,BRWV,tobias, kubis, satoshi} where the three-body unitarity was implemented differently by using Faddeev-like decomposition\cite{BR,tobias}, Kuri-Trieman equation\cite{kubis} and triangle diagrams \cite{satoshi}. Although   the inclusion 
of the three-body FSI is  necessary for a complete description 
of the decay process, it brings  new complex loop structures that increase significantly the  computation.

The main purpose of this work is the identification of the leading structures  acting in $\dkkk$ decay to be applied to the data as an alternative to isobar model and it 
 is organized as follows. The Triple-M amplitude is discussed 
in sections II to IV and the suggested amplitude for data fitting is given in section V.
Some simulations and general remarks are given in section VI. Details of the 
calculations are given in the appendices.

\section{model}

In the study of heavy meson decays, one deals with matrix
elements of both vector and axial currents between hadronic states.
Usually, the structure of these matrix elements  is parametrized in terms of  form-factors
and their shapes tend to be associated with resonances.

At low energies, one deals mostly with matrix elements
involving single particle states, which have been widely 
considered in the literature.
However, in the case of some leptonic reactions and heavy meson decays, 
available energies can be large enough for allowing the simultaneous 
production of several pseudoscalars.
Proper multi-meson structures become then relevant.
For instance, the process $e^-\,e^+ \rar 4\,\p$ involves the 
matrix element  $\la \p\p\p\p|J_\g^\m|0\ra$, $J_\g^\m$ being the
electromagnetic current  \cite{EU}.
A similar matrix element, with $J_\g^\m$ replaced with the
weak current $(V\sm A)^\m$, is instrumental to the description 
of the decay $\tau \rar \n \, 4\p$  \cite{EU}.

In the case of $D$ and $B$ hadronic decays, a rich structure 
 of multi-meson final states has been identified in the large 
amount of recent data.
Theoretical descriptions of these decays
involve two distinct sets of interactions.
The first one concerns the primary weak vertex, in which a heavy quark,
either $c$ or $b$, emits a $W$ and becomes a $SU(3)$ quark.
As this process happens within the heavy meson, 
it corresponds to the effective decay of a $D$ or a $B$ into 
a first set of $SU(3)$ mesons.
This is followed by purely hadronic final state interactions, 
in which the mesons produced in the weak decay rescatter, 
before being detected.
As both weak and final state interactions include several 
competing processes, the treatment of heavy meson decays 
into hadrons is necessarily involved.

One usually begins with the topologies given by Chau  \cite{Chau},
which implement CKM quark mixing for processes based on a single $W$.
Actual calculations, however, require the incorporation of Chau's scheme
into effective hadronic descriptions.
A possibility is to use factorization, as in the work of 
Bauer, Stich and Wirbel  \cite{BSW}.
Or alternatively, one can depart from effective Lagrangians \cite{HM}.

As instances, one mentions recent studies of the decay $D^+ \rar K^- \p^+ \p^+$.
Processes based on the axial current require the product 
of matrix elements
$\la \p^+ | A^\m |0\ra \la \p^+ K^- | A_\m|D^+ \ra$  \cite{Dkpp}, 
whereas those emphasizing the vector current
rely on the product
$\la \p^+\p^+ K^- | V^\m|\Kb^0\ra \la \Kb^0| V_\m |D^+\ra$,
with the $\Kb^0$ kept inside a loop  \cite{BR, BRWV}.   

In the present work, we concentrate on effects associated with the matrix element 
$\la K^-\, K^+ \, K^+ |A^\m|\, 0 \,\ra$, 
which is especially relevant for the decay $D^+ \rar K^-\,K^+\, K^+$.

\begin{figure}[h] 
\includegraphics[width=2.5cm,angle=90]{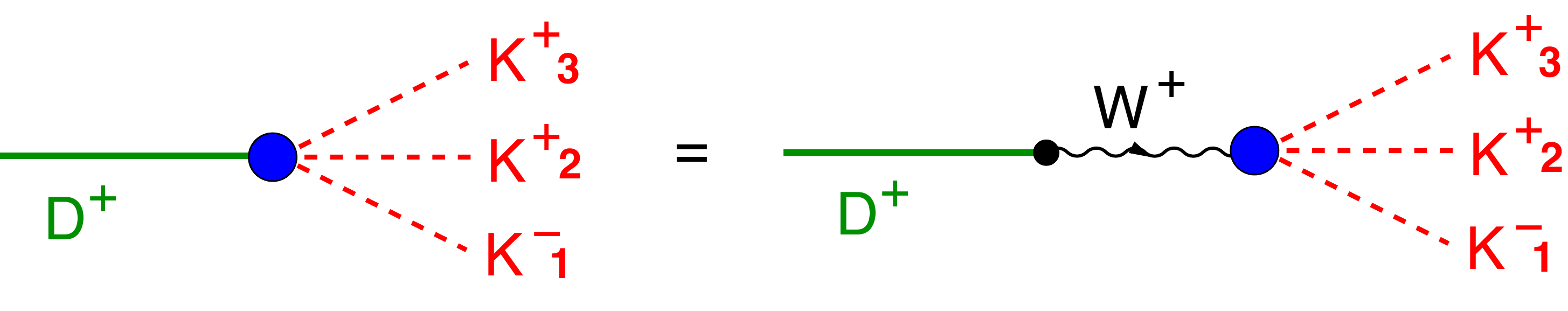}
\caption{The decay $D^+ \rar K^-\,K^+\, K^+$ (left)
is assumed to proceed thought quark-annihilation topology in the steps 
$D^+ \rar W^+$  and $W^+ \rar K^-\,K^+\, K^+$ (right).}
\label{FH}
\end{figure}

We  assume the decay to be dominated by the process shown in Fig.~\ref{FH}  
and determine the multi-meson matrix element by means of 
chiral effective theories.
The motivation for chiral symmetry is that the 
$u$-, $d$-, and $s$-quark masses are small when compared with the 
QCD scale $\Lambda \sim 1\,$GeV, indicating that the light sector 
of the theory is not far from the massless limit,
which is symmetric under the chiral $SU(3)\times SU(3)$ flavour group.
In this approach, light-quark condensates are included naturally
and pseudoscalar mesons are Goldstone bosons.
The {\em Chiral Perturbation Theory} was originally designed to 
describe low-energy interactions \cite{Wchi,GL}, where it yields 
the most reliable representation of 
the Standard Model.
Its scope was later enlarged, with the inclusion of resonances
as chiral corrections \cite{EGPR}, 
coupling schemes suited for heavy mesons \cite{HM},
and partial unitarization, by means of diagram
ressummations \cite{OO}.

Here, one is concerned just with the simplest possible structures,
which could be instrumental to empirical data analyses.
We work within the $K$-matrix approximation and,
therefore, skip loop contributions from off-shell states.
Our results are compatible with the conservation of the vector current (CVC)
and partial conservation of the axial current (PCAC). 
The latter is especially relevant for this problem, since it 
 implies that the divergence of the axial current
is proporcional to $M^2_K $.

\section{tree-level axial currents}
\label{tree}

The decay $D^+(P) \rar K^-(p_1)\,K^+(p_2)\, K^+(p_3)$ is assumed to proceed thought 
the intermediate steps 
$D^+ \rar W^+$  and $W^+ \rar K^-\,K^+\, K^+$, as in Fig. \ref{FH}.
The former  is associated with the matrix element
\bea
\la \,0\,| A^\m |\, D^+(P) \ra 
&\!=\!& -i\,\rtw\,F_D \,P^\m \;,
\label{ac.1}
\eea 
where $F_D$ is a constant and $P=(p_1 \sp p_2 \sp p_3)$. 

\begin{figure}[ht!] 
\includegraphics[width=8cm,angle=90]{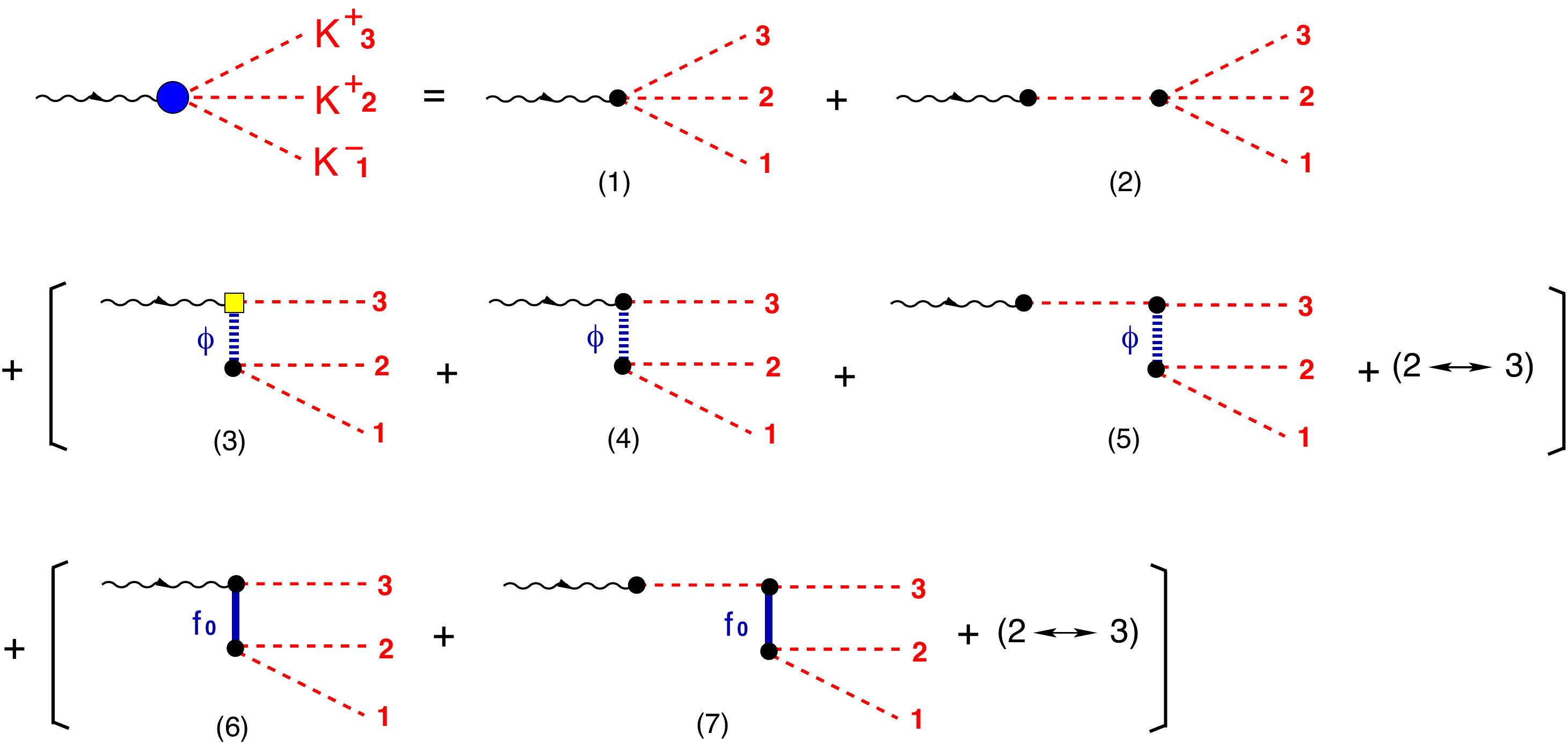}
\caption{Tree-level structure for the $A^\m \rar K^-\,K^+\, K^+$ matrix element:
the top line is LO and terms within brackets, which involve $\f$ and $f_0$
intermediate states, are NLO;
there are two different forms for the $WK\f$ coupling, 
indicated by a yellow box and a black dot.}
\label{FA3}
\end{figure}

This work relies heavily on the chiral effective lagrangians
including resonances, developed by Ecker, Gasser, Pich 
and De Rafael \cite{EGPR}, where  the formalism needed
can be found.
In particular, one follows their conventions for coupling constants.
The  tree-level matrix element
$\la K^-(p_1) \, K^-(p_2) \, K^+(p_3) |A^\m|\,0\,\ra$ is given in Fig. \ref{FA3}, 
where the top line displays the leading order (LO) contact terms, 
whereas  NLO corrections are given within brackets.
Individual contributions  are given in Appendix \ref{itc}.
The full contact term term $(1)+(2)$ includes a kaon pole and can be written as
\bea
&& \la K^-(p_1)\, K^+(p_2) \, K^+(p_3) |A^\m|\, 0 \,\ra_{(c)}
= - i\, \lb \frac{2\rtw}{ F_K}\rb \; \frac{1}{P^2 \sm M_K^2}
\nn\\[2mm]
&& \times 
\lc  \lb  P^2 \, (p_2 + p_3)^\m - P\cd(p_2 + p_3)  P^\m \rb + M_K^2 \, p_1^\m\rc \;,
\label{ac.2}
\eea
where $F_K$ is the kaon decay constant.

The treatment of the $\f$-meson 
includes  the $\p\p\p$ intermediate channel. 
This contribution is assumed to  be saturated by $\rho \p$  intermediate states
and is already included into the $\f$ propagators in Fig. \ref{FA3}.
The structure of this dressed propagator, 
denoted by $[ D_\f^{\p\rho}]^{-1}$, is discussed in Appendix \ref{dp}
and given by eq. ({\ref{dp.18}).
The $\phi$ contribution involves two independent terms, 
namely $(3)$ and $(4)+(5)$,  and reads
\bea
&& \la K^-(p_1) K^+(p_2) K^+(p_3) |A^\m|\,0\,\ra_{(\f)} 
\nn\\
&& = - i\, \lb \sin^2\theta\, \frac{3\, \rtw \,F_V \, G_V}{2\,F_K^3}\rb \; 
\frac{\lb P \cd p_2 \, p_1^\m - P \cd p_1 \, p_2^\m \rb }{D_\f^{\p\r}(m_{12}^2)} 
\nn\\[2mm]
&&+ i\, \lb \sin^2\theta\, \frac{3\, \rtw \,G_V^2}{F_K^3}\rb \; 
\frac{1}{P^2 \sm M_K^2} \;
\lc  \frac{1}{D_\f^{\p\r}(m_{12}^2)}
\right.
 \nn\\[2mm]
 && \times 
 \left.
\lb  p_2 \cd p_3 \, \lp P^2\,p_1^\m  - P\cd p_1 \, P^\m \rp  
-  p_1 \cd p_3 \,  \lp P^2 \, p_2^\m - P\cd p_2 \, P^\m \rp
\right.\right.
\nn\\[2mm]
&& \left. \left. -  M_K^2 \,  
\lp p_2 \cd p_3\,p_1^\m - p_1 \cd p_3 \, p_2^\m  \rp  \rb 
 \rc + (2  \lrar 3) \;,
\label{ac.3}
\eea
where $\theta$ is the $\o-\f$ mixing angle and $F_V$ and $G_V$ are  coupling constants 
defined in Ref. \cite{EGPR}.

As the role of the scalar $f_0(980)$ in the $SU(3)$ structure is not clear, we allow it to be either a singlet or a member of an octet (hereafter we refer to  $f_0(980)$ only as $f_0$).
In the formal developments, these two possibilities are labelled by $(0)$ and $(8)$,
respectively, and treated  together as long as possible. 
The terms $(6)+(7)$ of Fig. \ref{FA3} yield
\bea
&& \la K^-(p_1) K^+(p_2) K^+(p_3) |A^\m|\,0\,\ra_{(f_0)} 
\nn\\
&& = i\,\lb \g_n \,\frac{\rtw}{3\,F_K^3} \rb  \;
\frac{1}{P^2 - M_K^2} \, 
\lc \frac{1 }{m_{12}^2 - m_{f_0}^2} 
\right.
\nn\\[2mm]
&& \left. \times \lb c_d \, (P^2\,p_3^\m - P \cd p_3\,P^\m)  
- M_K^2 ( c_d\,p_3^\m - c_m \, P^\m )\rb \; 
\right.
\nn\\[2mm]
&& \left. \times 
\lb c_d \, m_{12}^2 - 2\, (c_d - c_m) \, M_K^2 \rb + (2  \lrar 3) \rc \;,
\label{ac.4}
\eea
where $c_d$ and $c_m$ are coupling constants (black dots in diagrams (6) and (7) in Fig.~\ref{FA3}) \cite{EGPR},	and we have used 
$\tilde{c}_i = c_i/\rth$ for the singlet, and $\g_0 = 8$  and $\g_8=1$.

The decay amplitude of the $D^+$, given in Fig. \ref{FH}, is 
\bea
T &\!=\!& \lb \frac{G_F}{\rtw} \, \sin^2\theta_C \rb \, \rtw\; F_D
\lb i \, P_\m\, \la K^-(p_1) K^+(p_2) K^+(p_3) |A^\m|\,0\,\ra \rb\;,
\label{ac.5}
\eea
where $\theta_C$ is the Cabibbo angle and $G_F$ is the Fermi weak constant.

Structures of the form $(P\cd y\, x^\m - P\cd x\, y^\m)$,
for two generic vectors $x^\m$ and $y^\m$, vanish when multiplied by
$P_\m$ and the  multi-meson current divergence is proportional 
to $M_K^2\,$, as expected by PCAC.
In terms of the form factors proposed by K\"uhn and Mirkes \cite{KM},
this means that the decay amplitude is proportional to their $F_4$.
Results (\ref{ac.2}-\ref{ac.4}) yield the tree contribution 
\bea
T_{tree}  &\!=\!& 
C\; \lc \lb M_D^2 + M_K^2 - m_{23}^2 \rb 
\right.
\nn\\[2mm]
&\!-\!& \left. \lb \sin^2\theta\, \frac{3 \,G_V^2}{4 F_K^2}\rb \; 
\lb \frac{m_{12}^2}{D_\f^{\p\r}(m_{12}^2)} \;\lp m_{13}^2 - m_{23}^2 \rp + 2\lrar 3 \rb  
\right.
 \nn\\[2mm]
&\!-\!& \left. \lb \frac{\g_n}{6\ F_K^2} \rb  \;
\lb \frac{1 }{m_{12}^2 - m_{f_0}^2} 
\lb c_d \,  \lp m_{12}^2 - M_K^2 \rp - (c_d - 2\, c_m )\, M_D^2 \rb \; 
\right.\right.
\nn\\[2mm]
 && \left. \left. \times  
\lb c_d \, m_{12}^2 - 2\,  (c_d - c_m) \, M_K^2 \rb + 2 \lrar 3 \rb \rc\;.
\label{ac.6}
\eea
\ni
where $C$ is the constant 
\bea
C &\!=\!&  \lb \frac{G_F}{\rtw} \, \sin^2\theta_C \rb \, \frac{2 F_D}{F_K} \; \frac{M_K^2}{M_D^2 - M_K^2} \;.
\label{ac.7}
\eea
The factor $C$ has dimension $[m]^{-2}$, given by the Fermi constant $G_F=1.166\times 10^{-5}\,$GeV$^{-2}$.

\section{final state interactions}
\label{fsi}

\begin{figure}[h] 
\includegraphics[width=0.6\columnwidth,angle=90]{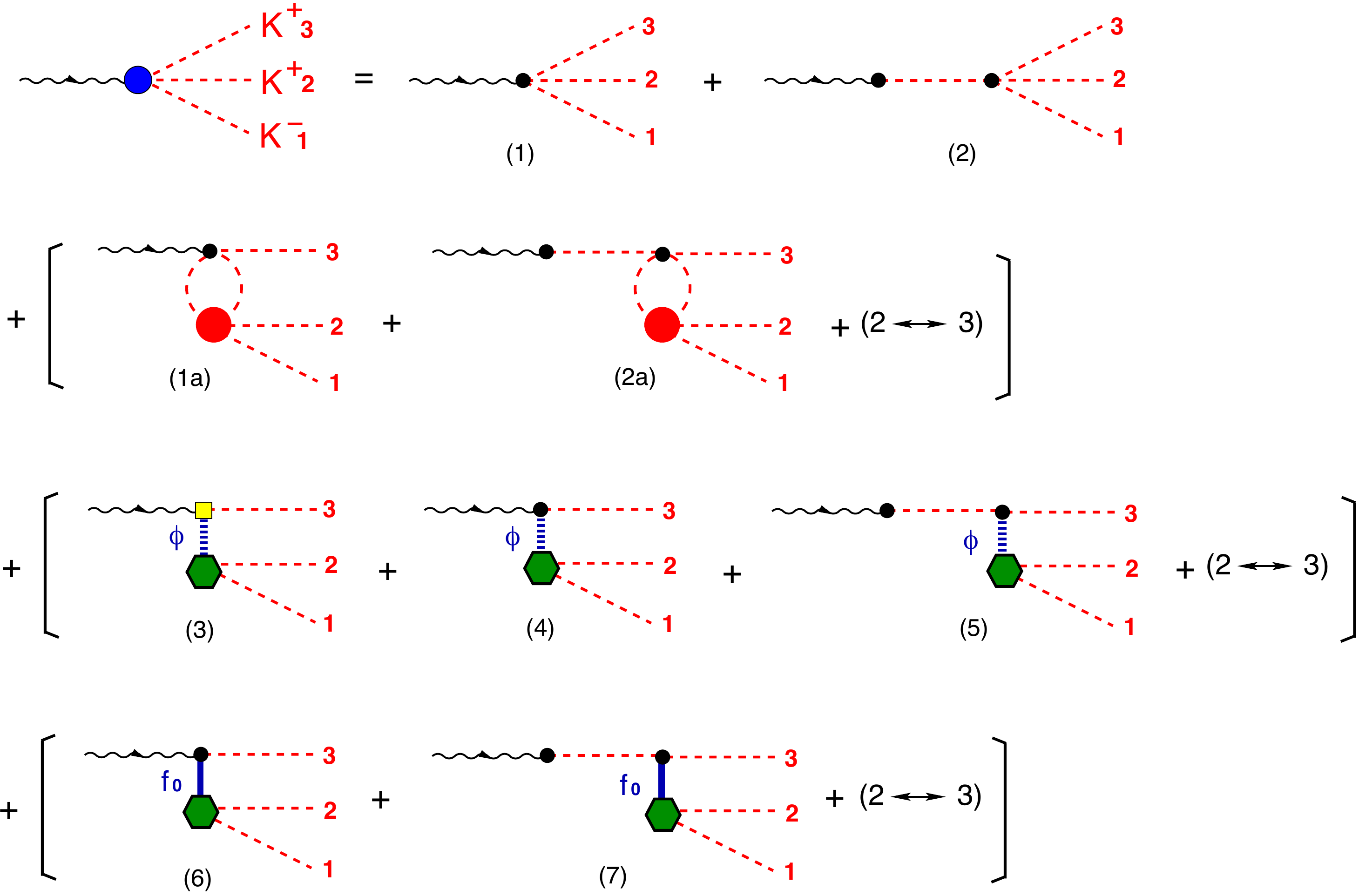}
\caption{Dynamical structure of the $A^\m \rar K^-\,K^+\, K^+$ matrix element, 
including final state interactions:
the top line, diagrams (1) and (2),  is the LO nonresonant contribution,
diagrams (1a) and (2a) include a $KK$ rescattering, indicated 
by the red blob;
whereas diagrams (3-7) describe  $\f$ and $f_0$ contributions
with their full widths;
there are two different forms for the $WK\f$ coupling, 
indicated by a yellow box and a black dot.}
\label{FSI}
\end{figure}

The full amplitude is obtained by including final state interactions in the processes
of Fig. \ref{FA3}.
Here we consider only two-body rescattering process, leaving the three-body  treatment to a future work. The corresponding dynamics is described in Fig. \ref{FSI} and involves two classes of contributions.
The first one, shown by diagrams (1a) and (2a), concerns $\Kb K$ rescattering.
In principle, the intermediate states could have isospin $I=0,1$ and angular 
momentum $J=0,1$. 
We keep only contributions with $I=0, J=1$, associated with the $\f$
channel,  and postpone the discussion of the other ones.
The second class corresponds to employing the production amplitudes,
derived in Appendix \ref{pa}, which endow the $\f$ and $f_0$ propagators with their full widths.

The nonresonant contribution is determined at tree-level. 
Dynamically, it is a proper three-body amplitude, a direct consequence of chiral symmetry,
which predicts multi-meson topologies.
This goes beyond  the notion of  a spectator particle,
as in the 2+1 approximation  (two-body subsystem+spectator).
It is given by a real polynomial, which can be written in two alternative forms
\bea
T_{NR}  &\!=\!& C \; \lb  \,  M_D^2 + M_K^2 - m_{23}^2 \,\rb \;,
\label{x.01}\\[2mm]
 &\!=\!& C \; \lb  \,  ( m_{12}^2 - M_K^2 )\,  +  2\lrar 3 \, \rb \;.
\label{x.02}
\eea
The second form makes it clear that $T_{NR}$ contains just $S$-waves.

The rescattering amplitudes, given by diagrams $(1a+2a)$  in Fig. \ref{FSI},  involve intermediate states 
with the same quantum numbers as the $f_0$ and the $\f$, which 
are denoted by $T_{Rf}$ and $T_{R\f}$.
We postpone the discussion of the former and 
next, we consider the latter, together with the $\f$ contribution $T_\f$, 
associated with diagrams $(3+4+5)$ and based on results from Appendices \ref{kk} and \ref{pa}.
They read
\bea
T_{R\f}  &\!=\!& - i\, C\, \lb \sin^2\theta\, \frac{3 \,G_V^2}{4 F_K^2} 
- \frac{3\,D_\f^{\p\r}}{8\,m_{12}^2}\rb \,
\lb \frac{\, m_{12}\,[Q_c^3+Q_n^3]}{16\p\,F_K^2} \;
\frac{[ m_{13}^2 - m_{23}^2 ]}{D_\f(m_{12}^2)} + 2\lrar 3 \, \rb   \;,
\label{x.1}\\[4mm]
T_\f  &\!=\!& - C\, \lb \sin^2\theta\, \frac{3 \,G_V^2}{4 F_K^2}\rb \,
\lb  m_{12}^2 \; 
\frac{[ m_{13}^2 - m_{23}^2 ]}{D_\f(m_{12}^2)} + 2\lrar 3 \, \rb  \;,
\label{x.2}
\eea
where
\bea
&& D_\f(s) =  s - m_\f^2  + i\, m_\f \, \G_\f(s) 
- \lb i\, \frac{1}{8\p\, F_K^2} \, \frac{D_\f^{\p\r}(s)}{\sqrt{s}}\; \lp  Q_{c}^3+ Q_{n}^3\rp \rb \;,
\label{x.3}\\[4mm]
&& m_\f \, \G_\f (s) = 
 \sqrt{s}\, \lb \Gamma_{KK} \; \frac{( Q_{c}^3 + Q_{n}^3)} {( \Qt_{c}^3 + \Qt_{n}^3)}  
+\Gamma_{\p\rho} \;\frac{s}{m_\f^2} \; \frac{Q_{\p \rho}^3}{\Qt_{\p \rho}^3} \rb \;,
\label{x.4}\\[4mm]
&& D_\f^{\p\r} (s) = s - m_\f^2  + i \,\G_{\p\r} \; \frac{s^{3/2}}{m_\f^2} \; \frac{Q_{\p\r}^3}{\Qt_{\p\r}^3} \;.
\label{x.5}
\eea
with 
$ Q_{\p\r} = \frac{1}{2} \sqrt{ s - 2\,  (M_\p^2 + \mr2 ) + (M_\p^2 - \mr2 ) ^2/s}\;$,
$Q_{c} = \frac{1}{2} \sqrt{s-4 M_{K^+}^2}\;$, $Q_{n} = \frac{1}{2} \sqrt{s-4 M_{K^0}^2}\;$
and $\Qt = Q(s=m_\f^2)$.
In Eq.(\ref{x.4}), $\G_{\p\rho}$ and $\G_{KK}$ are the decay width of the $\f$ and   their values can be found in PDG \cite{PDG}. 
The use made of  $\G_{KK}$ fixes the  $\phi$ coupling constant,
as in eq.(\ref{c.33}),   to be 
\bea
&& \sin^2\th \, \frac{3\, G_V^2}{4\, F_K^2} = \frac{3\p\,F_K^2 \,\Gamma_{KK}}{( \Qt_{c}^3+\Qt_{n}^3)  } \;.
\label{x.6}
\eea 

The functions $T_{R\f}$ and $T_\f$ share the same dressed $\f$ propagator, 
owing to the presence of the  $\Kb K$ amplitude in the former.

\begin{figure}[h] 
\includegraphics[width=0.46\columnwidth,angle=0]{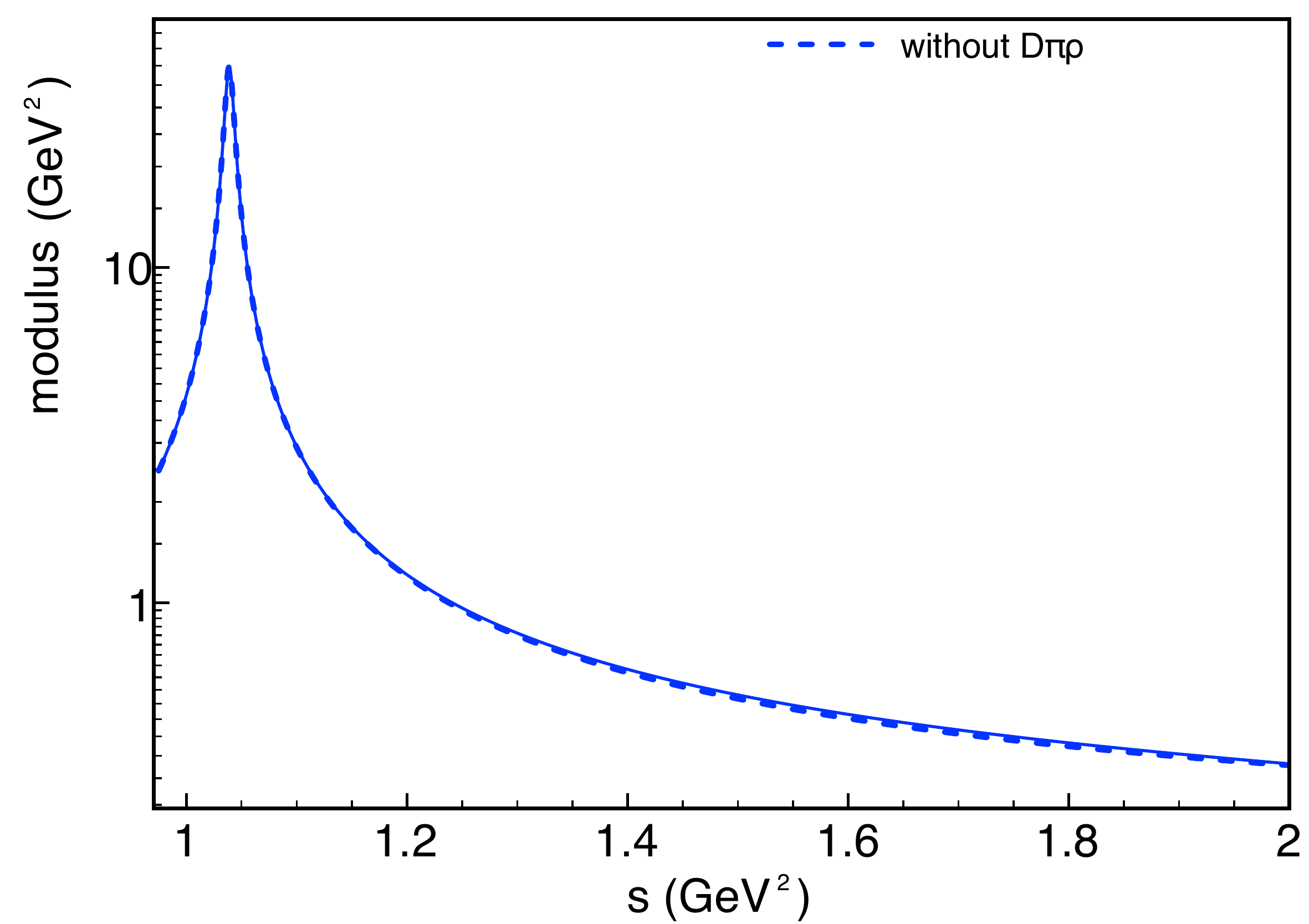}
\hspace*{5mm}
\includegraphics[width=0.46\columnwidth,angle=0]{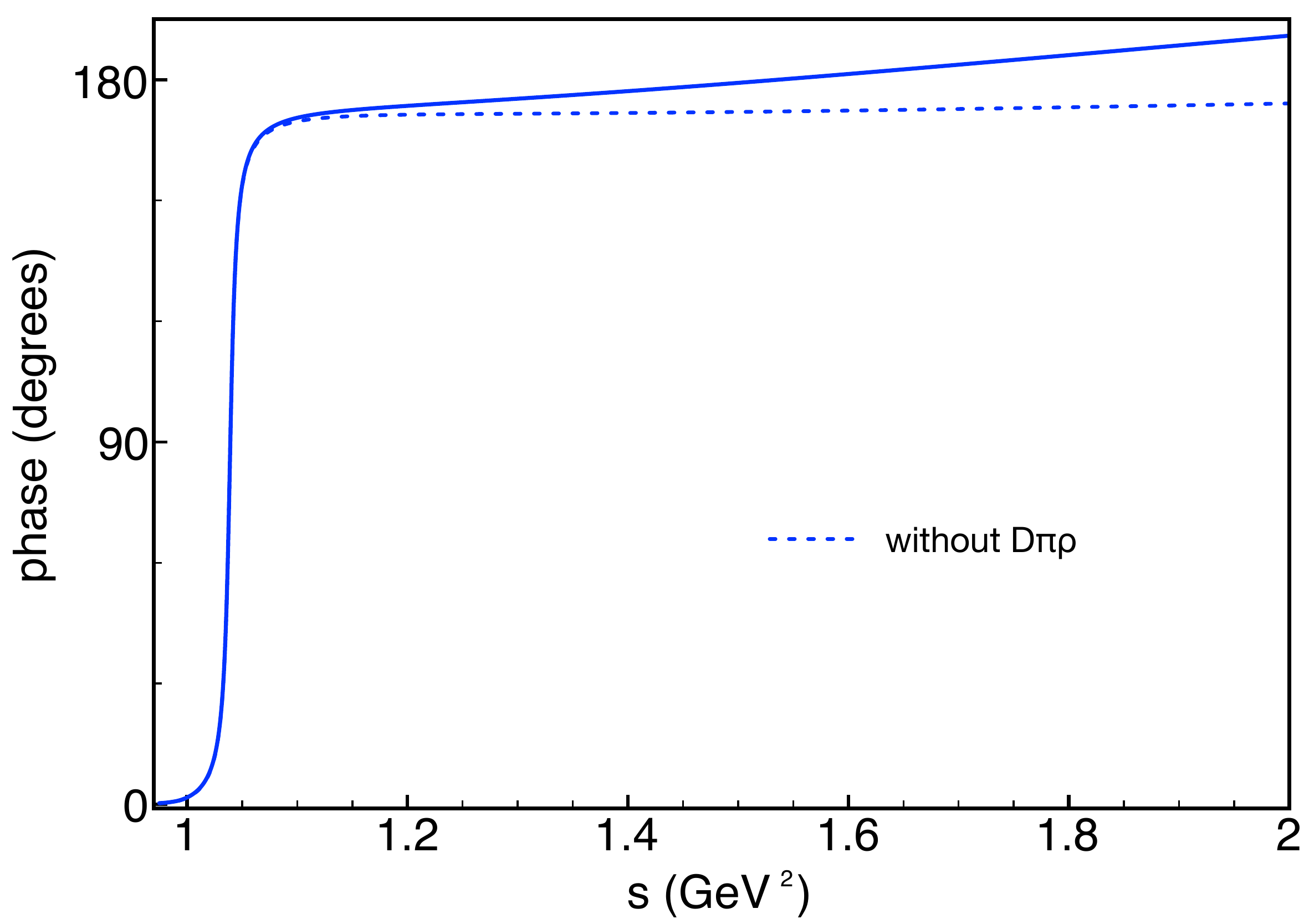}
\caption{ Moduli (left) and phases (right)  of the ratio $[T_{R\f}+T_\f]/[-C\,(m_{13}^2 - m_{23}^2)]$,
eqs.(\ref{x.1}) and (\ref{x.2}), 
with  (continuous blue) and without (dotted blue)
the factor $D_\f^{\p\r}$, eq.(\ref{x.5}).}
\label{Dpirho}
\end{figure}

Our results contain several new features. 
One of them regards the description of intermediate $\Kb K$ interactions,
which rely on contributions from both
resonances and pont-like processes, as discussed in Appendix \ref{kk}.
The latter vanish at the peak of the resonance and increase as one moves away from it.
Their signature are the terms proportional to  $D_\f^{\p\r}$.
In Fig. \ref{Dpirho}, we assess the importance of intermediate contact interactions,
by comparing results for the dimensionless ratio  
$[T_{R\f}+T_\f]/[-C\,(m_{13}^2 - m_{23}^2)]$,
with and without $D_\f^{\p\r}$.
One learns that contact interactions 
have little numerical relevance in the  phase-space accessible to the $\dkkk$ decay.

In Fig. \ref{TR} we display the relative importance of  the ratios 
$T_{R\f}/[-C\,(m_{13}^2 - m_{23}^2)]$ and $T_\f/[-C\,(m_{13}^2 - m_{23}^2)]$,
and notices that the latter is largely dominant.
The dip on the curve associated with $T_{R\f}$, around $s=1.8\,$GeV, is due to the destructive 
interference between the two terms in the first bracket of eq.(\ref{x.1}) in that region.

\begin{figure}[h] 
\includegraphics[width=0.46\columnwidth,angle=0]{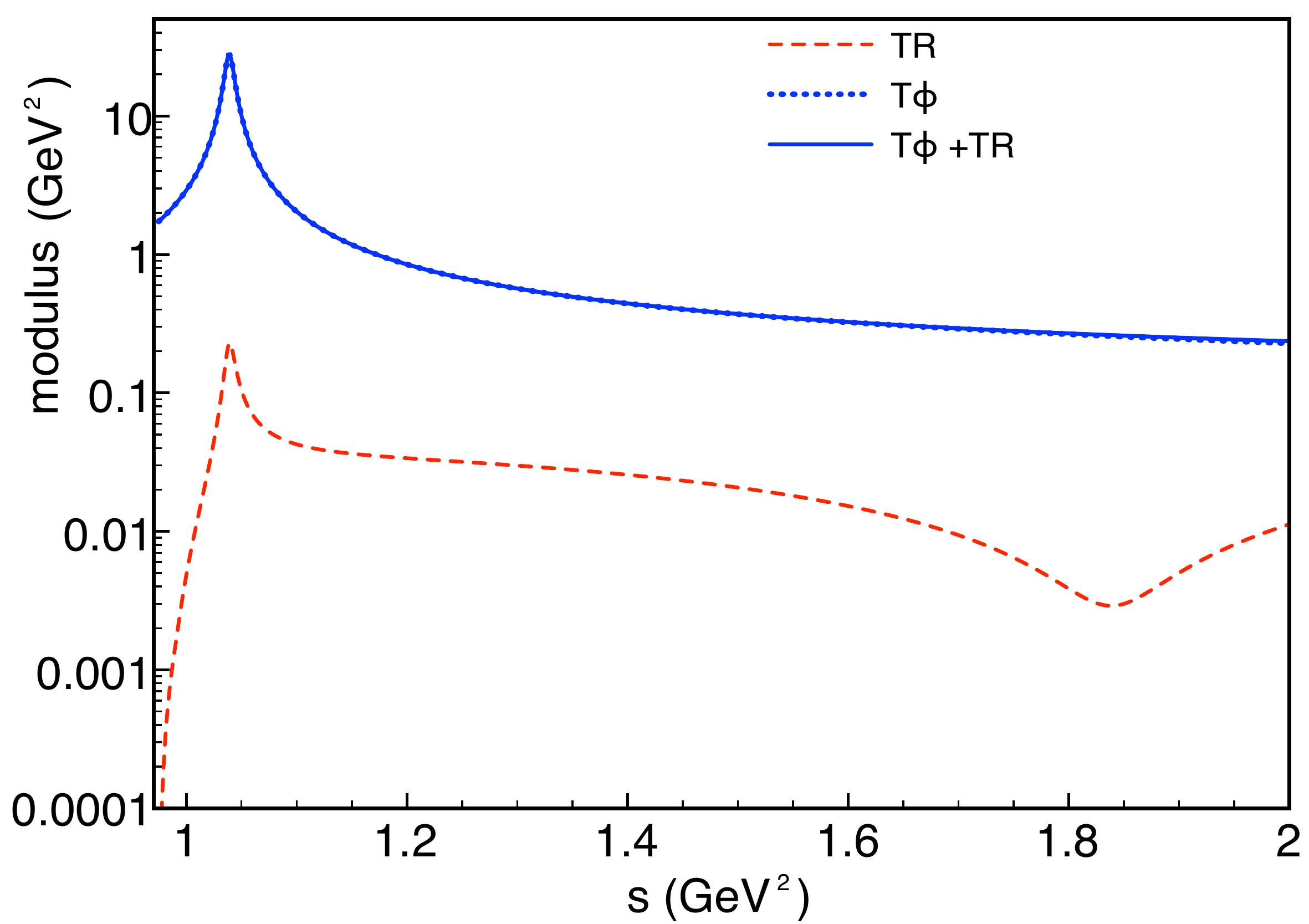}
\hspace*{5mm}
\includegraphics[width=0.46\columnwidth,angle=0]{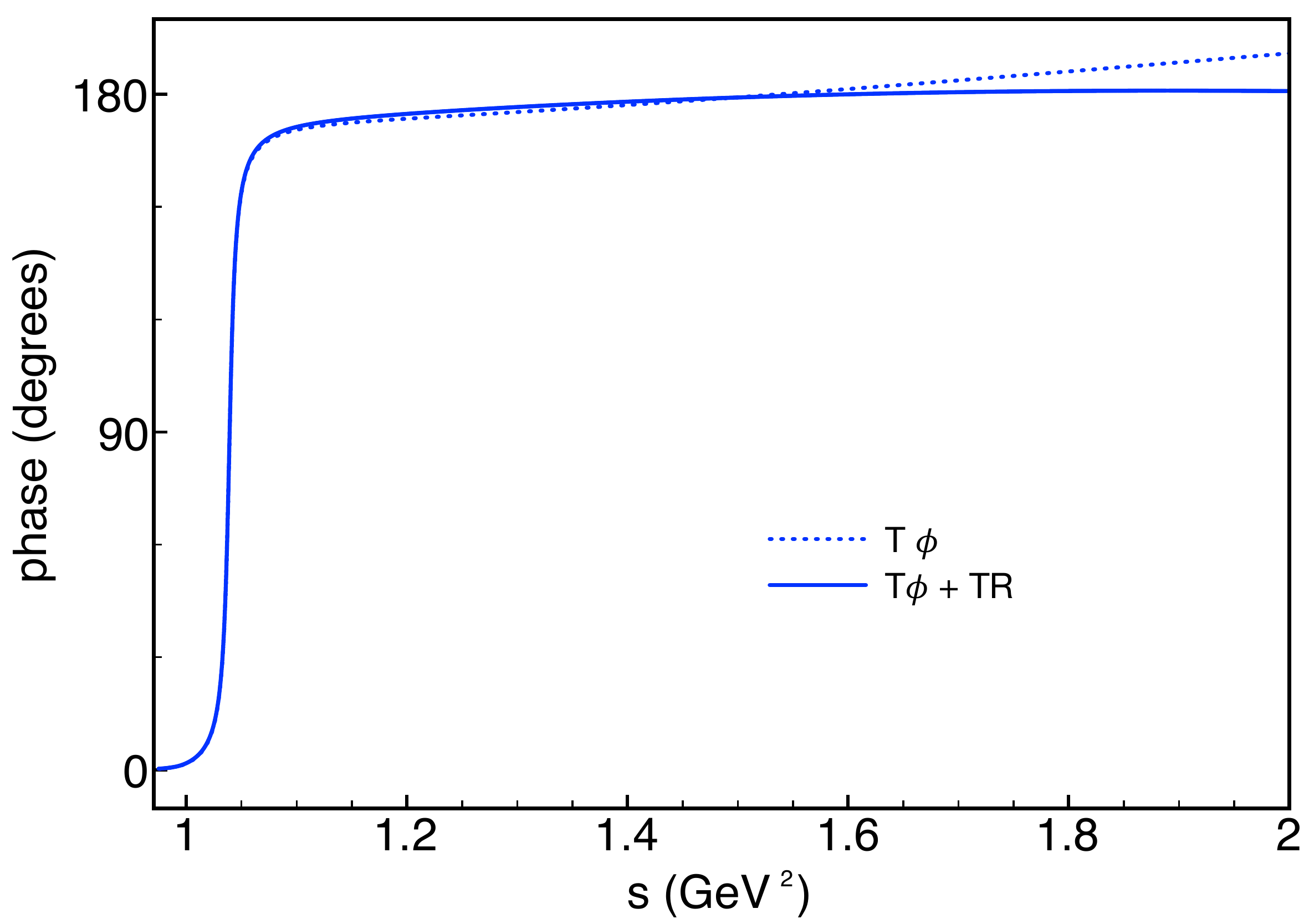}
\caption{ Moduli(left) and phases(right)  of 
$T_{R\f}/[-C\,(m_{13}^2 - m_{23}^2)]$, eq.(\ref{x.1}) (dashed red), 
$T_\phi/[-C\,(m_{13}^2 - m_{23}^2)]$, eq(\ref{x.2}) (dotted blue) 
and $[T_\f + T_R]/[-C\,(m_{13}^2 - m_{23}^2)]$ (continuous blue).}
\label{TR}
\end{figure}

The preceding discussion indicates that, in the $\f$ channel,  
both $D_\f^{\p\r}$ and $T_{R\f}$ contributions are small and,
for the sake of simplicity,  they can be safely removed from the model.
Another new feature in our results 
concerns the independent widths for $K^- K^+$, $\Kb^0 K^0$, and $\rho \p$ decay modes, 
which give rise to the structures proportional to $Q_c$,  $Q_n$ and $Q_{\p\r}$
in eq.(\ref{x.4}).
A  last issue is  the factor $m_{12}^2$ in the numerator of eq.(\ref{x.2}), 
and the $\sqrt{s}$ outside the bracket in eq.(\ref{x.4}),
which are signatures of resonance couplings in chiral perturbation theory. 
Thus, the  leading contribution in the $\f$-channel is proportional to 
the dimensionless function
\bea
A_{TM\f} = \frac{s}
{ s - m_\f^2  + i\, m_\f \; \G_{TM\f}}\;,
\label{x.7}
\eea
with $m_\f \; \Gamma_{TM\f}$ given by eq.(\ref{x.4}).
In Fig. \ref{BW},  we compare it with the  usual relativistic Breit-Wigner (BW) \cite{asner} for the same channel,
employed in most Dalitz plot analyses,  excluding barrier and spin factors, which reads
\bea
 && A_{BW\f} = \frac{m_\f^2}
{ s - m_\f^2  + i\, m_\f \; \G_{BW\f }}
\label{x.8}
\\[2mm]
&& m_\f \, \G_{BW\f}(s)= \G_\f \lb \frac{m_\f^2}{\sqrt{s}}\rb \;  \frac{ Q_{c}^3(s)}{\Qt_{c}^3} \;,
\\[2mm]
&& \G_\f = \G_{KK} + \G_{\p\r} \;,
\eea
The factor $m_\phi^2$ in the numerator of eq.(\ref{x.8}) was introduced so that it
has the same normalization as eq.(\ref{x.7}), at the pole.  
The main differences between both expressions are the factors proportional to $s/m_\f^2$, 
associated with chiral symmetry.
Their effects are already visible at  $s\sim 1.2\,$GeV and  increase with energy.
In particular it is interesting to note that the modulus of the BW falls faster than that of the Triple-M
at high values of $s$.
%
%
%
\begin{figure}[ht!] 
\hspace*{-3mm}\includegraphics[width=0.46\columnwidth,angle=0]{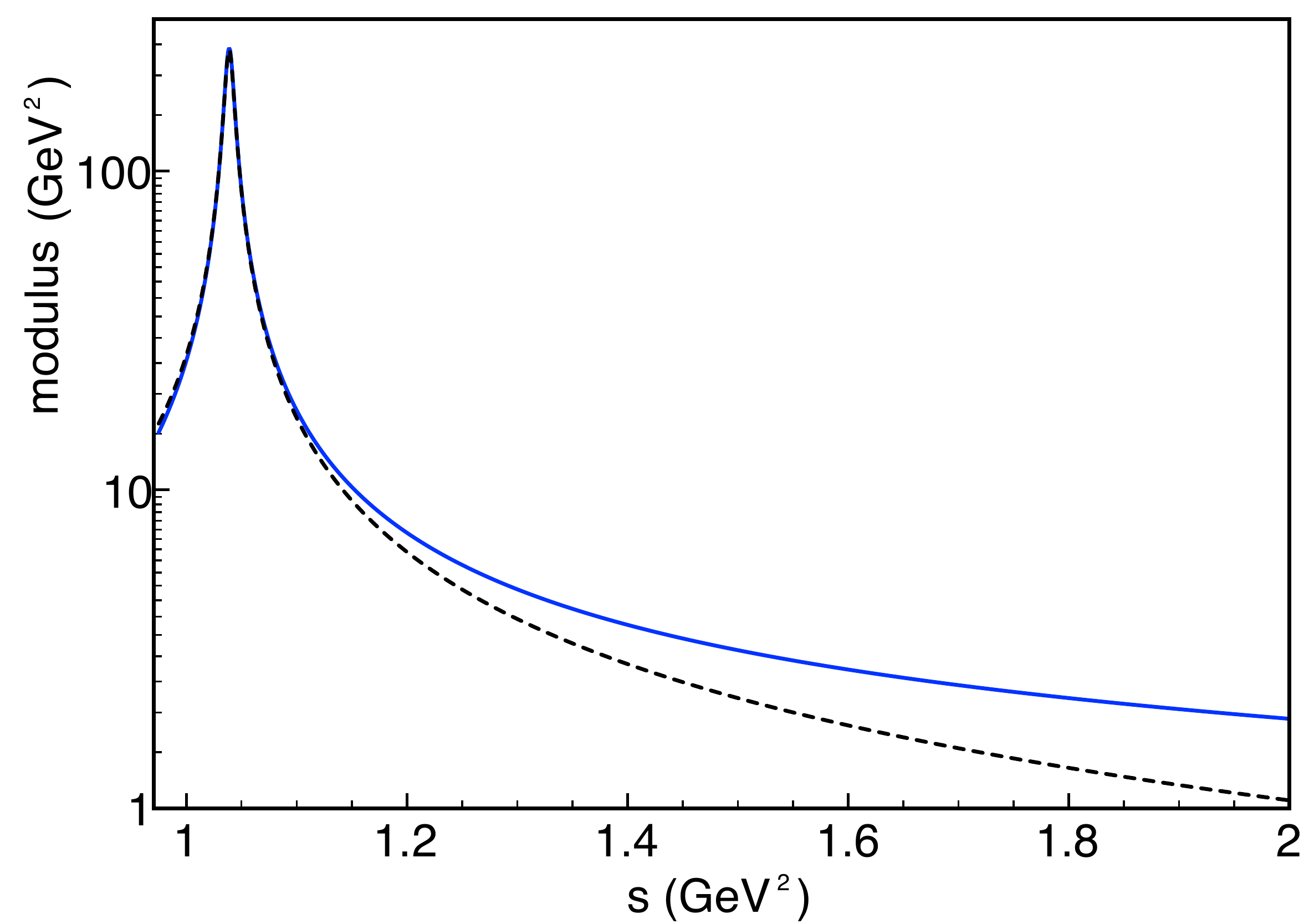}
\hspace*{5mm}
\includegraphics[width=0.46\columnwidth,angle=0]{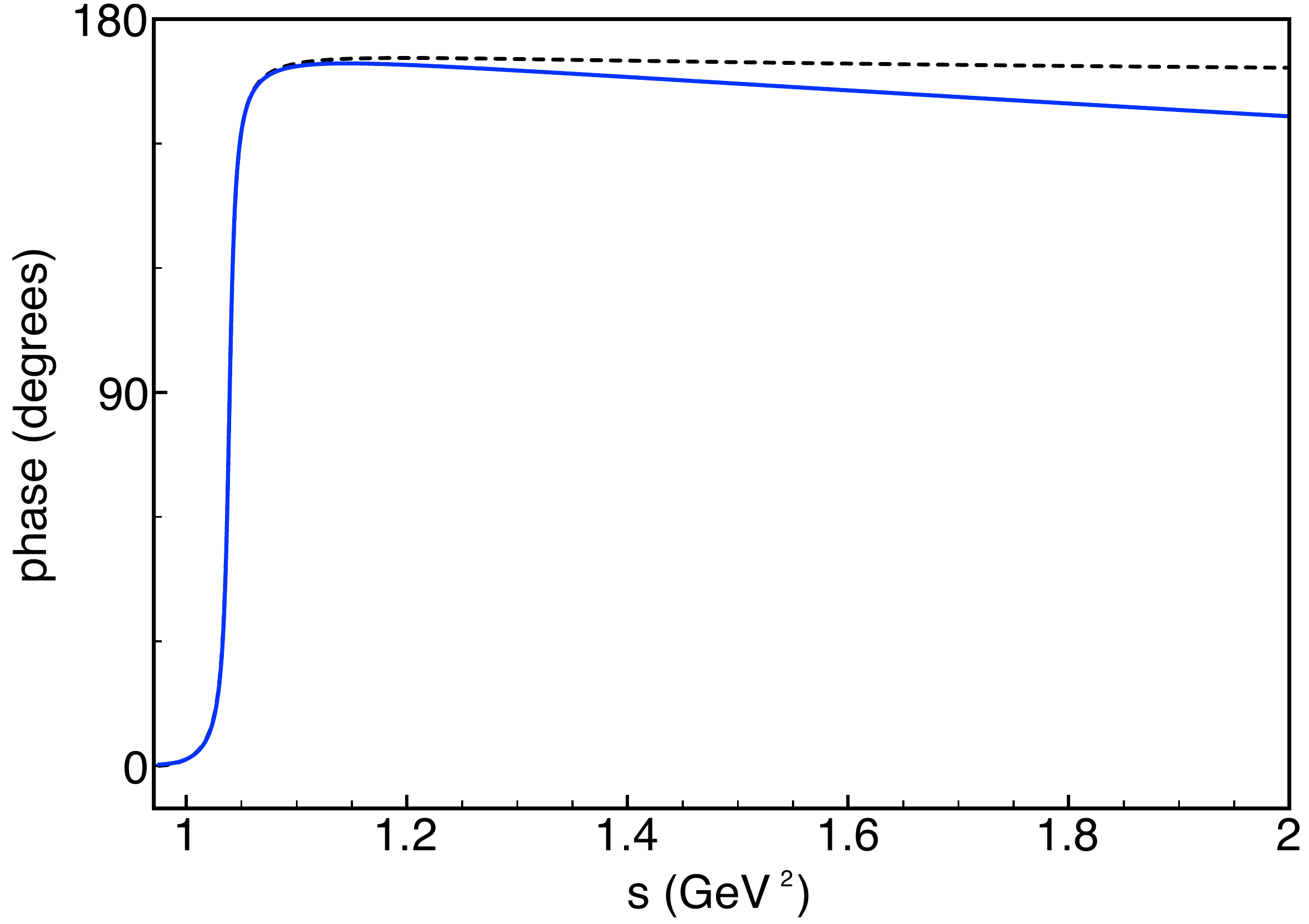}
\caption{Moduli (left) and phases (right)  of the functions 
$A_{TM\f}$, eq.(\ref{x.7}),  (continuous blue) and 
$A_{BW\f}$, eq.(\ref{x.8}), (dashed black).}
\label{BW}
\end{figure}

The $SU(3)$ status of the $f_0$ is uncertain and, for instance, one could follow Ref. \cite{EGPR},
assuming it to be a linear combination of singlet and octet states.
However, owing to the exploratory nature of this work, 
we consider it to be either a singlet or an octet, labelled by subscripts $(0)$ and $(8)$.  
These possibilities affect both the intensity of the $f_0$ coupling and the relative proportion 
of $\p\p$ and $\Kb K$ in its decay modes. 
The two alternatives are written as 
\bea
T_{f_0} &\!=\!& - C \lb \frac{\g_n }{6 F_K} \rb  \;
\lb \lb c_d \,  \lp m_{12}^2 - M_K^2 \rp - (c_d - 2\, c_m) \, M_D^2 \rb \; 
\frac{G_K(m_{12}^2)}{D_n(m_{12}^2)} + 2\lrar 3 \, \rb \;,
\label{x.9}
\eea
with $\g_0=8\,$, $\g_8=1\,$, and  
\bea
&& D_n(s) =  s-  m_{f_0}^2  + i\, m_{f_0} \, \G_n(s) \;,
\label{x.10}\\[2mm]
&&  m_{f_0}\,\G_0 = \frac{G_\p^2}{4\p F_\p^2} \, \frac{Q_{\p\p}}{\sqrt{s}}
+ \frac{G_K^2}{3\p F_K^2} \, \frac{Q_{KK}}{\sqrt{s}} 
+ \frac{G_\eta^2}{12\p F_\eta^2} \, \frac{Q_{\eta\eta}}{\sqrt{s}} \,\Theta[s - 4\,M^2_\eta]  \;,
\label{x.11}\\[2mm]
&&  m_{f_0}\,\G_8 = \frac{G_\p^2}{8\p F_\p^2} \, \frac{Q_{\p\p}}{\sqrt{s}}
+ \frac{G_K^2}{24\p F_K^2} \, \frac{Q_{KK}}{\sqrt{s}} 
+ \frac{G_\eta^{2}}{24\p F_\eta^2} \, \frac{Q_{\eta\eta}}{\sqrt{s}}\,\Theta[s - 4\,M^2_\eta]  \;,
\label{x.12}
\eea
\bea
&& G_P(s) = \frac{\lb c_d\,s - 2\,(c_d \sm c_m)\,M_P^2 \rb}{F_P}   \;,
\label{x.13}\\[2mm]
&& Q_{PP} (s) =  \frac{1}{2} \sqrt{s-4\,M_{P}^2}\;,
\label{x.14}
\eea 
for $P=\p, K, \eta$.

\begin{figure}[ht!!] 
\includegraphics[width=0.46\columnwidth,angle=0]{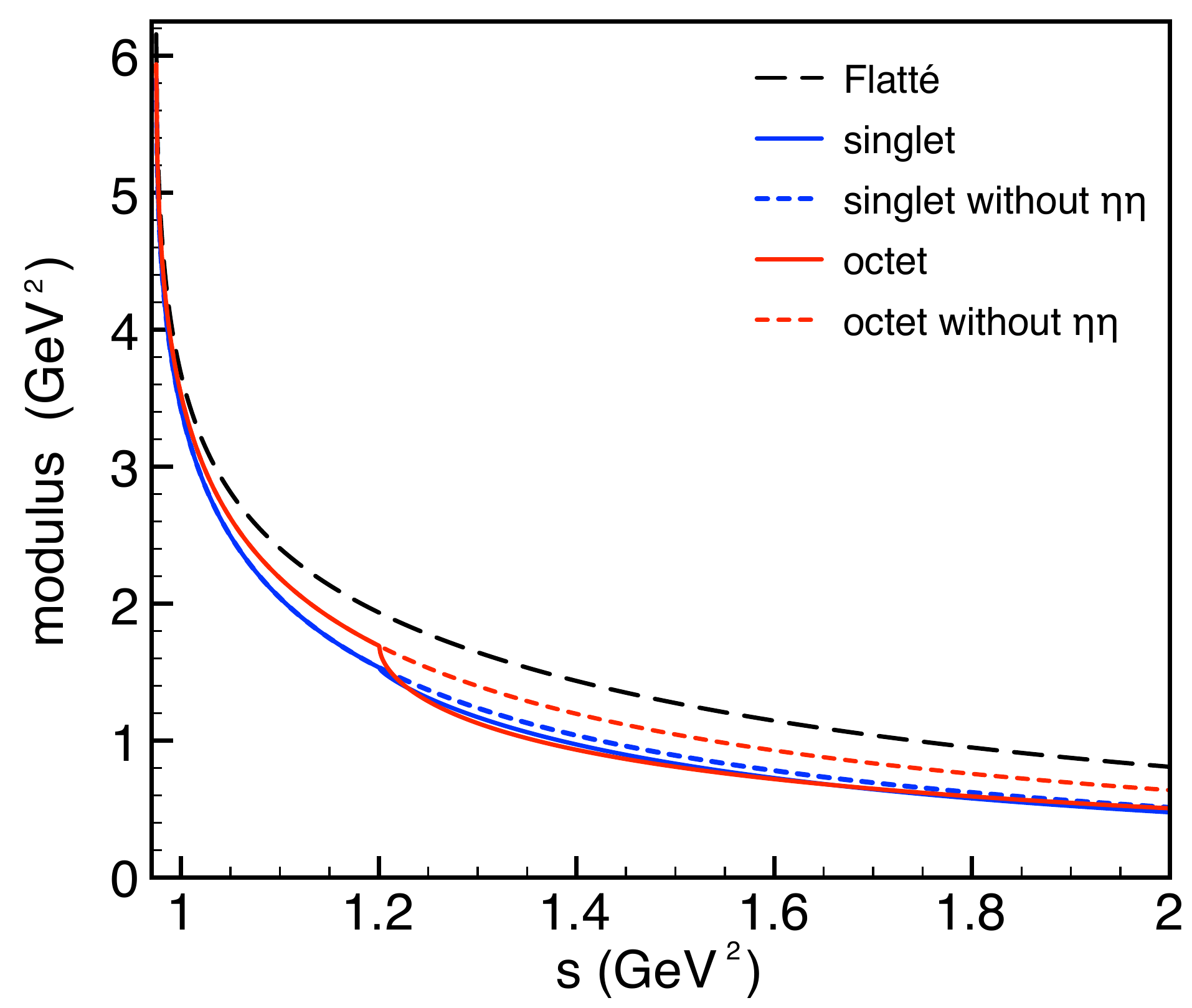}
\hspace*{5mm}
\includegraphics[width=0.46\columnwidth,angle=0]{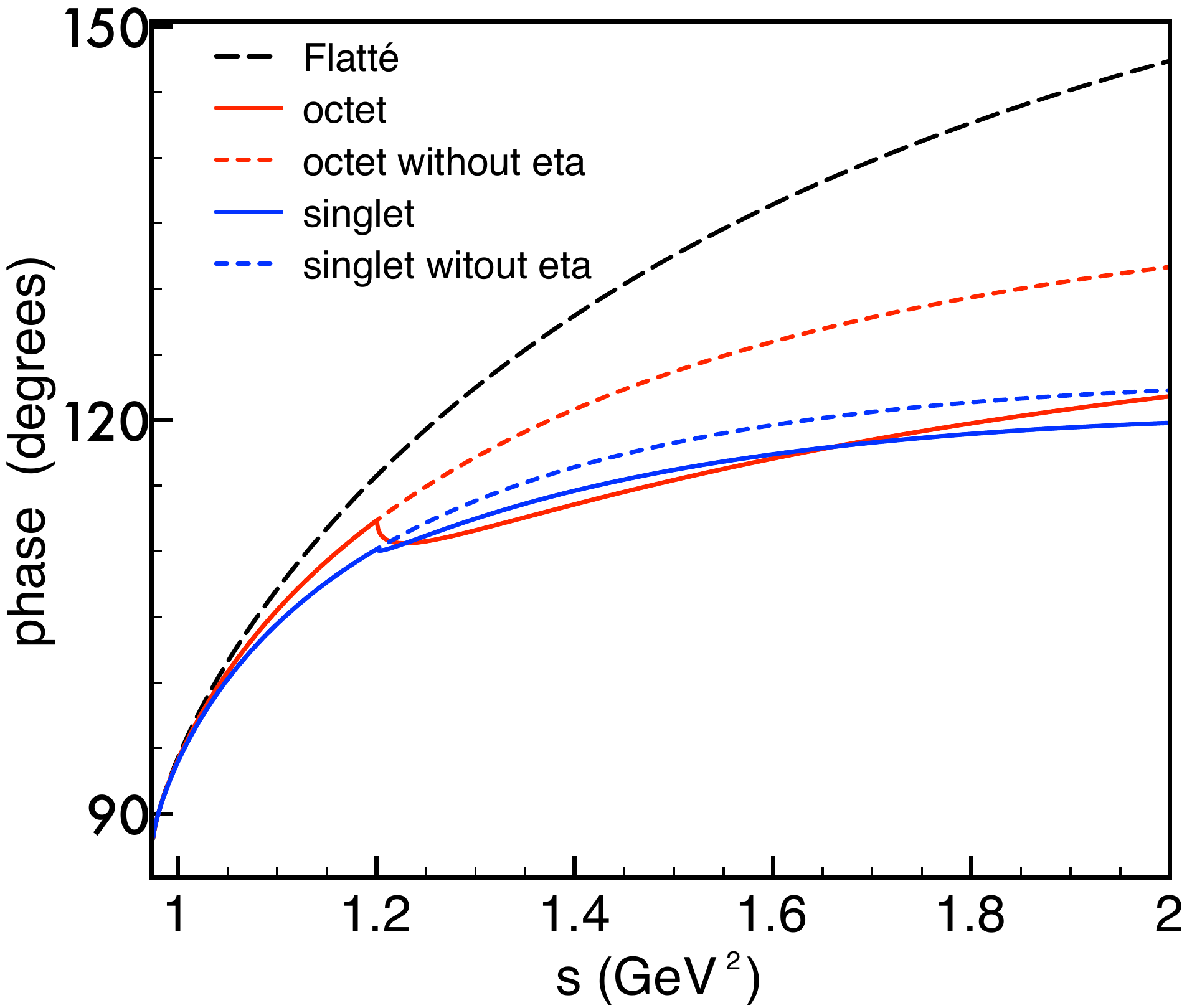}
\caption{Moduli(left) and phases(right)  of the dimensionless amplitudes
$A_{Flatte}$, eq.(\ref{x.16}) (dashed black)  and 
$A_{TM\,f_0}$, eq.(\ref{x.15}) for the singlet (continuous blue)
and octet (continuous red) cases;
the cusps in the last two curves are due to the opening of the $\eta\eta$ channel;
the elimination of this coupling gives rise to the corresponding dashed curves.}
\label{Flat}
\end{figure}

For the sake of completeness, we have allowed the $f_0$ to couple to an intermediate 
$\eta\eta$ state, even if it becomes relevant at higher energies only.
The most striking feature of the $f_0$ in the Triple-M is the presence of $s$-dependent couplings, 
predicted by chiral perturbation theory \cite{EGPR}.
This means that the amplitude $T_{f_0}$ is somewhat  flexible, 
since it depends on two free coupling parameters, namely $c_d$ and $c_m$.
In Fig. \ref{Flat},  we compare the dimensionless functions 
\bea
A_{TM\,f_0}= \frac{m_{f_0}^2}{s- m_{f_0}^2  + i\, m_{f_0} \, \G_n(s) } \;,
\label{x.15}
\eea
for $n=0,\,8\,$, eqs.(\ref{x.11}) and (\ref{x.12}),
with  a Flatt\'e function \cite{Fla},  with parameters obtained by BES from  $J/\psi \to \f \p\p (K\,K)$ data [$g_{\pi\pi}=0.165\,$GeV and 
$g_{KK}=0.695\,$GeV ] \cite{bes}
\bea
A_{Flatte}=\frac{m_{f_0}^2}
{ s - m_{f_0}^2 + i\,m_{f_0}\, 2\,
\lb g_{\pi\pi}\, Q_{\p\p} +\,g_{KK}\, Q_{KK} \rb/\sqrt{s} }.
\label{x.16}
\eea

In order to make the comparison meaningful, we fix the parameters $c_d$ and $c_m$ in eq.(\ref{x.15})  
so that $A_{TM f_0}$ and $A_{Flatte}$ have the same modulus  at $s=m_{f_0}^2$. This yields  $[c_d, \,c_m = 0.016, \, 0.069\,]\,$GeV, for the singlet,
and $[c_d ,\, c_m = 0.018, \, 0.220\,] \,$GeV, for the octet.
Inspecting  figure 7, one notes that effect of the $\eta\eta$ channel 
manifest  as threshold cusps around $s\sim 1.2\,$GeV
and remain visible afterwards.
Even when the $\eta\eta$-coupling is neglected, the differences between the
Flatt\'e function and Triple-M predictions remain important.

The dependence of $T_{f_0}$, eq.(\ref{x.9}), on the free parameters $c_d$ and $c_m$
is a rather strong one.
This is illustrated in Fig. \ref{EGPR}, where we display our singlet and octet predictions
for the ratio $T_{f_0}/[-C]$, based on the choices used  in Fig. \ref{Flat},
without the $\eta\eta$-coupling, together with those obtained by using the values  proposed by
Ecker, Gasser, Pich e Rafael\cite{EGPR}, namely 
$[c_d , \, c_m = 0.032,  \, 0.042\,] \,$GeV.
It is clear that the  values of $c_d$ and $c_m$  can affect considerably the final line shape 
of the $f_0$ and
one is entitled to expect fits to data to be quite sensitive to these parameters.

\begin{figure}[h!!] 
\includegraphics[width=0.46\columnwidth,angle=0]{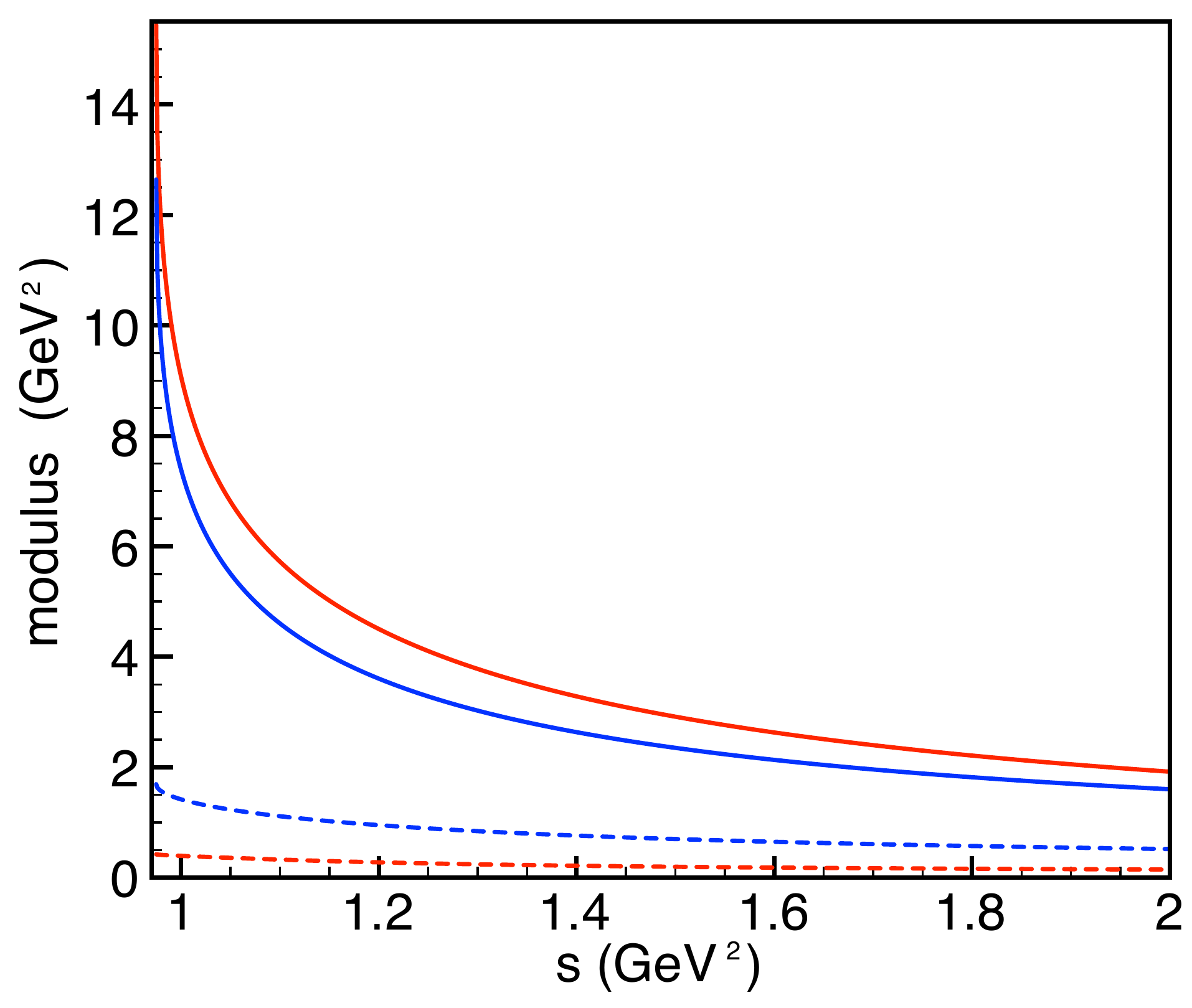}
\hspace*{5mm}
\includegraphics[width=0.46\columnwidth,angle=0]{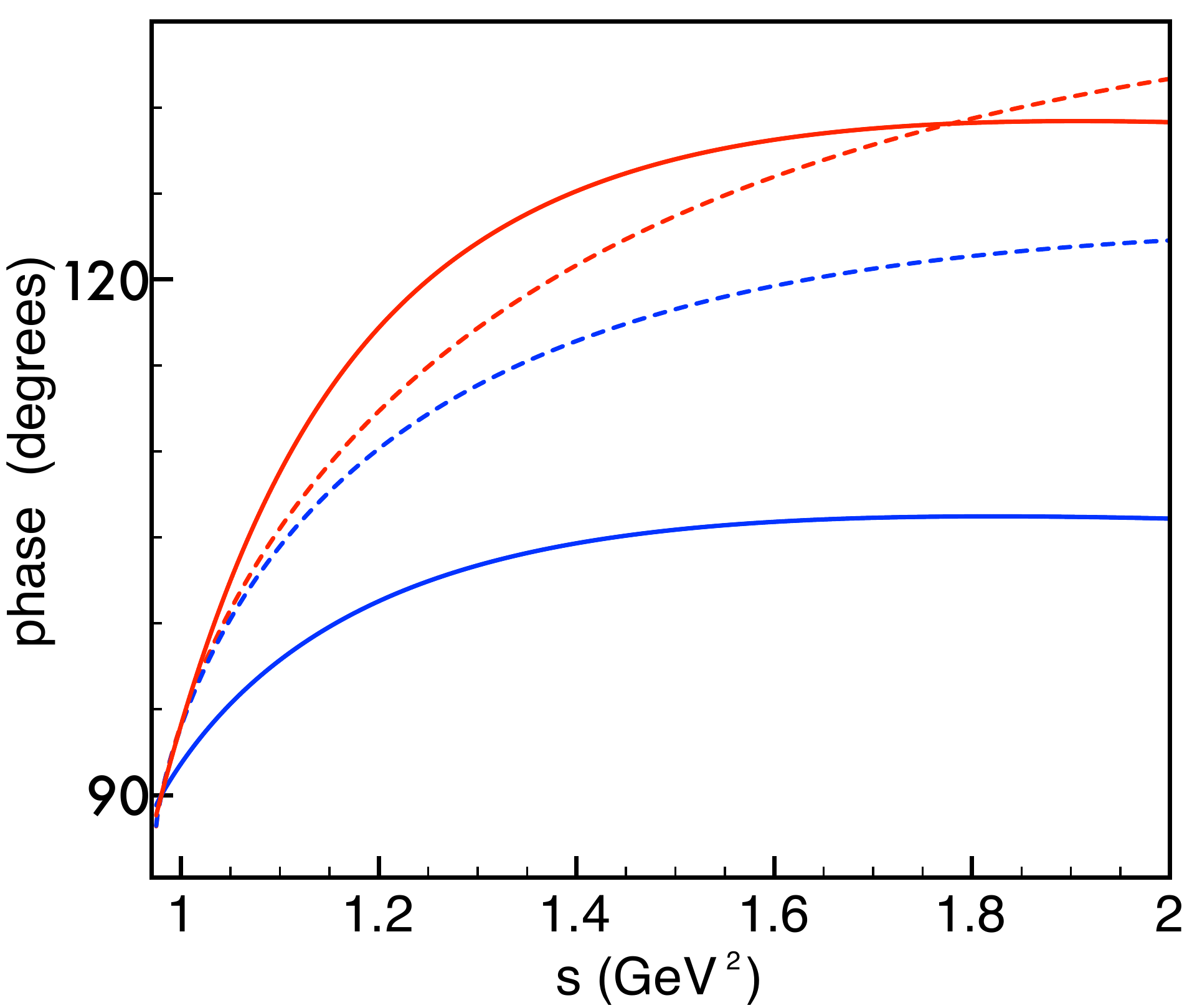}
\caption{Moduli(left) and phases(right)  of the ratio $T_{f_0}/[-C]$, eq.(\ref{x.9}) 
for the singlet (blue curves) and octet (red curves) cases, 
based on the same values of $c_d$ and $c_m$ as in Fig. \ref{Flat} 
(continuous curves) and on those given in Ref. \cite{EGPR} (dashed curves).}
\label{EGPR}
\end{figure}

%

In the Triple-M, the $S$-wave receives contributions from both $T_{NR}$, 
and $T_{f_0}$.
Their relative importance is assessed in Fig. \ref{SWave}, where we 
display the ratios $T_{NR}/[-C]\,$, eq.(\ref{x.02}),
$T_{f_0}/[-C]\,$,  eq.(\ref{x.9}), and their sum,
using the parameters $c_d$ and $c_m$  fixed by the Flatt\'e function, eq.(\ref{x.16}). 
The figure indicates that the interference of these two terms is mostly destructive.
The $f_0$ dominates at low energies whereas the nonresonant
interaction increase linearly with the energy.
The resulting profile for the modulus of  the $S$-wave 
falls from threshold up to $s\sim2\,$GeV, 
where it has a minimum.
This pattern of interference is also  important for the phase.

\begin{figure}[ht!!] 
\includegraphics[width=0.46\columnwidth,angle=0]{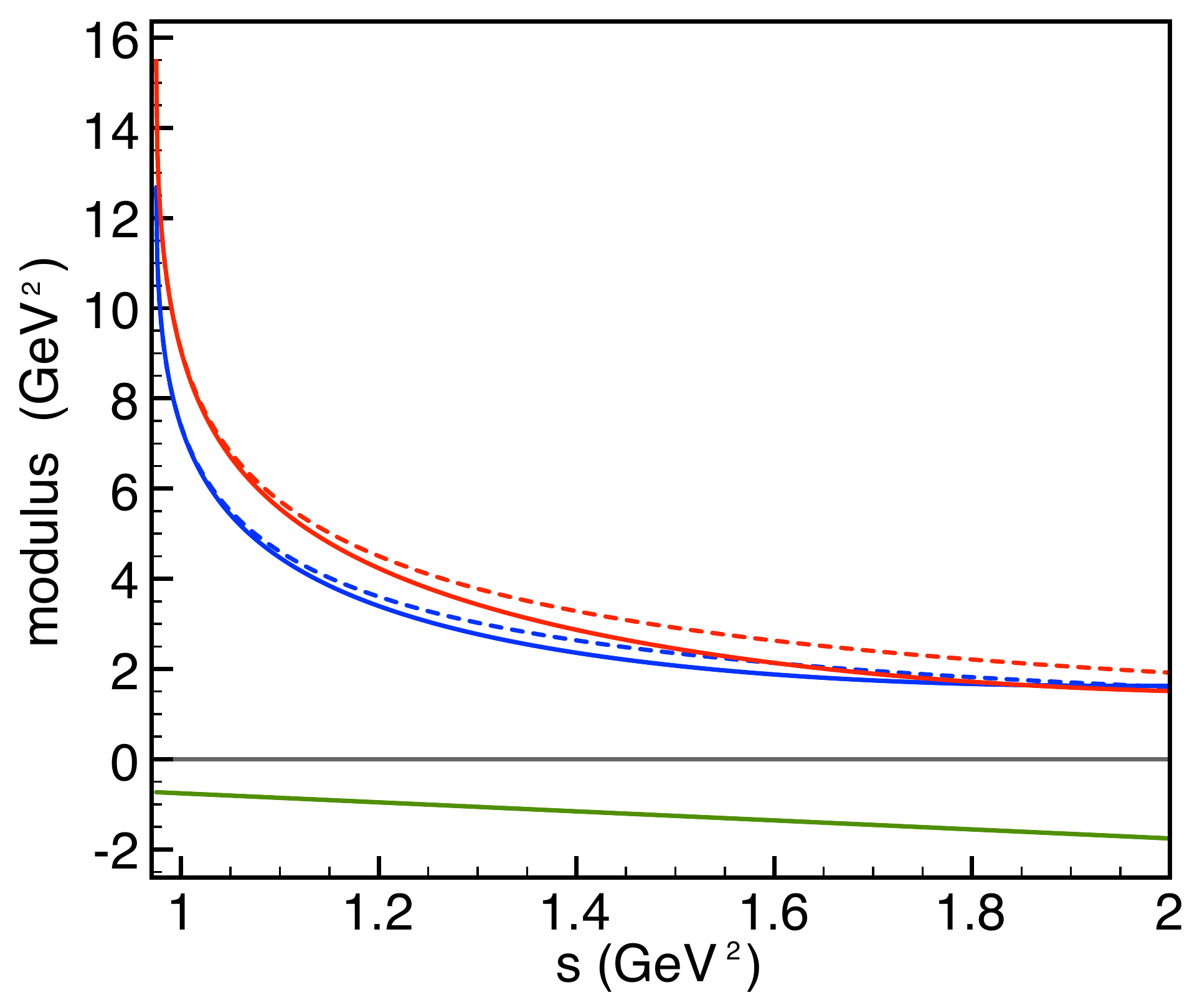}
\hspace*{5mm}
\includegraphics[width=0.46\columnwidth,angle=0]{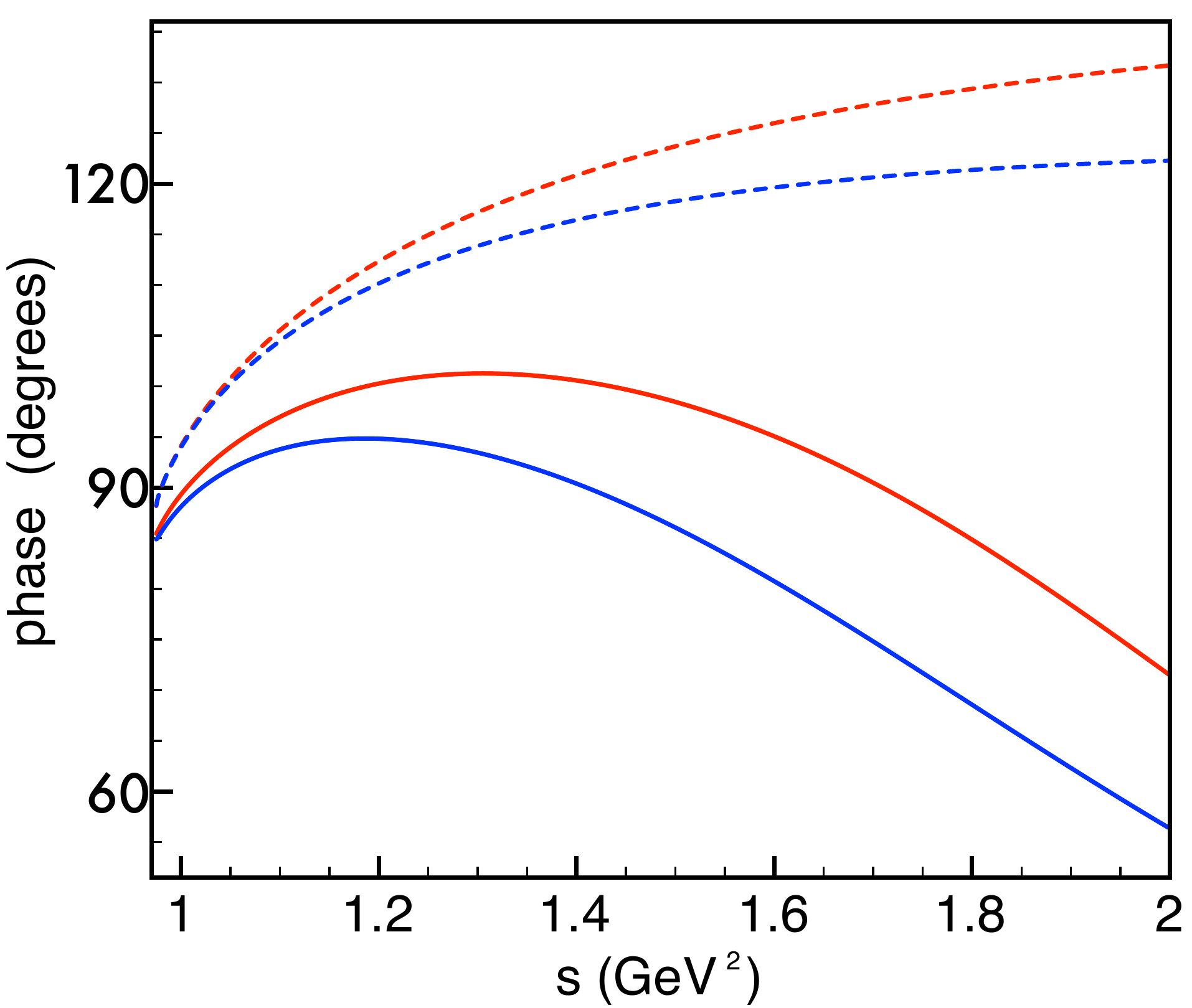}
\caption{Moduli(left) and phases(right)  for the dimensionless ratios
$T_{NR}/[-C]\,$, eq.(\ref{x.02}), (continuous green),
$T_{f_0}/[-C]\,$,  eq.(\ref{x.9}), for the singlet (dashed blue)
and octet (dashed red) cases,
together with their sums (continuous blue and red).}
\label{SWave}
\end{figure}

%

\section{The Multi-Meson Model - Triple-M}
\label{MMM}

The main purpose of this work is to provide a reliable model to be used for fitting data.
Our full results incorporate the dynamics  described in Fig. \ref{FSI} and 
the corresponding mathematical expressions were given in the previous section, where the leading contributions of each kind were identified.
Here, we present a simplified version of the Triple-M amplitude, composed by these leading 
contributions, which contain  three free parameters, associated to the scalar $f_0$.
Our Triple-M amplitude include  a nonresonant contribution ($NR$), 
supplemented by $\f$ and $f_0$ resonant terms,
and is formally written as%
\bea
T_{_{M\!M\!M}} &\!=\!& T_{NR} + \,T_\f + \, T_{f_0} \;,
\label{m.1}
\eea
The nonresonant term is given directly from diagrams $(1+2)$ in  Fig. \ref{FSI}  and reads
\bea
T_{NR}  &\!=\!& C \; \lc \,  M_D^2 + M_K^2 - m_{23}^2 \,\rc \;,
\label{t.2}
\eea
where $C$ is the constant with dimension $[m]^{-2}$, given in eq.(\ref{ac.7}) 
\bea
C &\!=\!&  \lb \frac{G_F}{\rtw} \, \sin^2\theta_C \rb \, 2\; \frac{F_D}{F_K} \; \frac{M_K^2}{M_D^2 - M_K^2} \;.
\nn
\eea
The amplitude $T_{R\f}$, eq.(\ref{x.1}), is neglected and the $\phi$
contribution is  
\bea
T_\f  &\!=\!& - C\, \lb \sin^2\theta\, \frac{3 \,G_V^2}{4 F_K^2}\rb \,
\lb  m_{12}^2 \; 
\frac{[ m_{13}^2 - m_{23}^2 ]}{D_\f(m_{12}^2)} + 2\lrar 3 \, \rb  \;,
\label{m.2}
\eea
with
\bea
&& D_\f(s) =  s - m_\f^2  + i\, m_\f \, \G_\f(s)  \;,
\label{m.3}\\[4mm]
&& m_\f \, \G_\f (s) = 
 \sqrt{s}\, \lb \Gamma_{KK} \; \frac{( Q_{c}^3 + Q_{n}^3)} {( \Qt_{c}^3 + \Qt_{n}^3)}  
+\Gamma_{\p\rho} \;\frac{s}{m_\f^2} \; \frac{Q_{\p \rho}^3}{\Qt_{\p \rho}^3} \rb \;.
\label{m.4}
\eea
There are no free parameters in $T_\f$.
The partial widths are $\G_{KK}= 3.55\,$MeV and  $\G_{\p\r}=0.65\,$MeV \cite{PDG}.
Using the former in eq.(\ref{x.6}), together with $F_K=0.107\,$GeV \cite{JOP},
$\Qt_c=126.41\,$MeV, and
$\Qt_n=110.10\,$MeV, one determines the coefficient of eq.(\ref{m.2}) as
\bea
&& \sin^2\th \, \frac{3\, G_V^2}{4\, F_K^2} = \frac{3\p\,F_K^2 \,\Gamma_{KK}}{( \Qt_{c}^3+\Qt_{n}^3)  }
= 0.1140 \;.
\nn
\eea
This expression also allows one to find the coupling constant $G_V$.
The standard value $\sin\theta=0.76$ \cite{PDG} yields $G_V= 0.055\,$GeV, which 
is quite close to the prescriptions from chiral perturbation theory \cite{EGPR} ($0.053 \,-\, 0.069 \,$GeV).

The amplitude $T_{f_0}$ is obtained by neglecting $\eta\eta$ couplings:  

\bea
T_{f_0} &\!=\!& - C \lb \frac{\g_n }{6 F_K} \rb  \;
\lb \lb c_d \,  \lp m_{12}^2 - M_K^2 \rp - (c_d - 2\, c_m) \, M_D^2 \rb \; 
\frac{G_K(m_{12}^2)}{D_n(m_{12}^2)} + 2\lrar 3 \, \rb \;,
\label{m.5}\\[2mm]
&& D_n(s) =  s-  m_{f_0}^2  + i\, m_{f_0} \, \G_n(s) \;,
\label{m.6}\\[2mm]
&&  m_{f_0}\,\G_0 = \frac{G_\p^2}{4\p F_\p^2} \, \frac{Q_{\p\p}}{\sqrt{s}}
+ \frac{G_K^2}{3\p F_K^2} \, \frac{Q_{KK}}{\sqrt{s}} \;,
\label{m.7}\\[2mm]
&&  m_{f_0}\,\G_8 = \frac{G_\p^2}{8\p F_\p^2} \, \frac{Q_{\p\p}}{\sqrt{s}}
+ \frac{G_K^2}{24\p F_K^2} \, \frac{Q_{KK}}{\sqrt{s}}  \;.
\label{m.8}
\eea
The functions $G_P$, eq.(\ref{x.13}), depend on the parameters $F_P$ and,
in the literature, one finds $F_\pi=0.093\,$GeV \cite{GL}, $F_K=0.107\,$GeV \cite{JOP}.
The values of $c_d$ and $c_m$ are to be determined by fits to data and 
the low-energy estimates $c_d, \, c_m = 0.032, \, 0.042\,$GeV \cite{EGPR}
provide educated points of departure.
For the propose of Monte-Carlo simulation, in the next section, we use the values obtained
 by comparing the $f_0$ width with Flatt\'e function, as discussed in the previous section.

\section{MC simulations}
In this section, simulations of the Triple-M amplitude are presented for the decay
$D^+\to K^-(p_1)K^+(p_2)K^+(p_3)$. Note that the convention is that the 
odd-charged particle is always  particle 1. Since the like-charged kaons are 
identical particles, the Dalitz plot is symmetric. This means 
$s_{12}\equiv(p_1+p_2)^2$ 
and $s_{13}\equiv(p_1+p_3)^2$ are equivalent. The third invariant,
$s_{23}\equiv(p_2+p_3)^2$, is the invariant mass squared of the
two like-charged kaons. The Dalitz plots are represented in terms of
$s_{12}$ and $s_{13}$ invariants. In this representation the $s_{23}$ axis
runs along the diagonal, with threshold ($4M_K^2$) at the upper border
of the Dalitz plot.

\begin{figure}[ht] 
\includegraphics[width=0.8\textwidth]{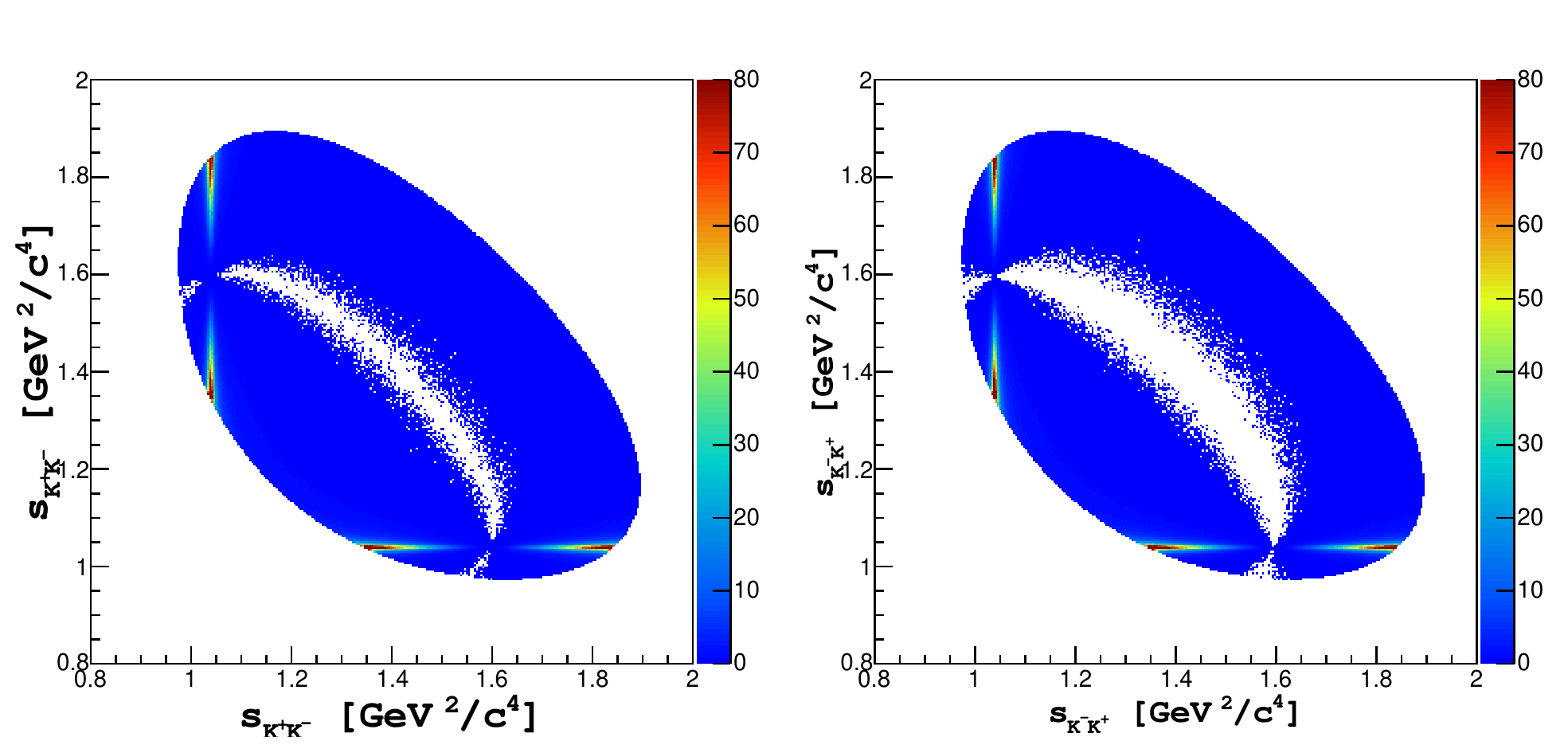}
\caption{Dalitz plot with the MC simulation of: a relativistic Breit-Wigner amplitude (left);
the $\phi$ component of the Triple-M  (right).} 
\label{tphi}
\end{figure}

\begin{figure}[ht] 
\includegraphics[width=0.8\textwidth]{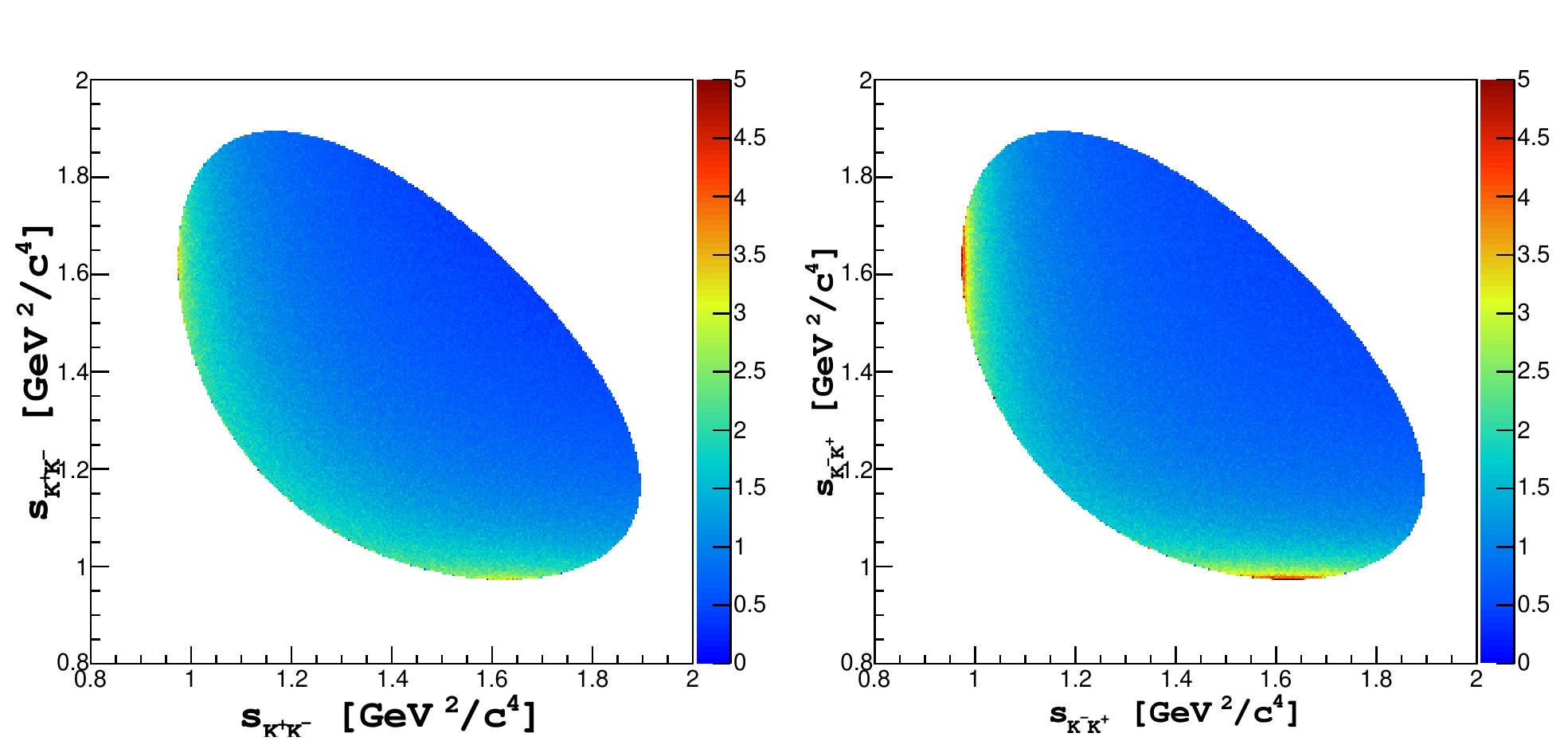}
\caption{Dalitz plot with the MC simulation of: the Flatt\'e amplitude (left);
 the $f_0$ singlet component of the Triple-M (right).}
\label{tf0}
\end{figure}

\begin{figure}[ht] 
\includegraphics[width=0.8\textwidth]{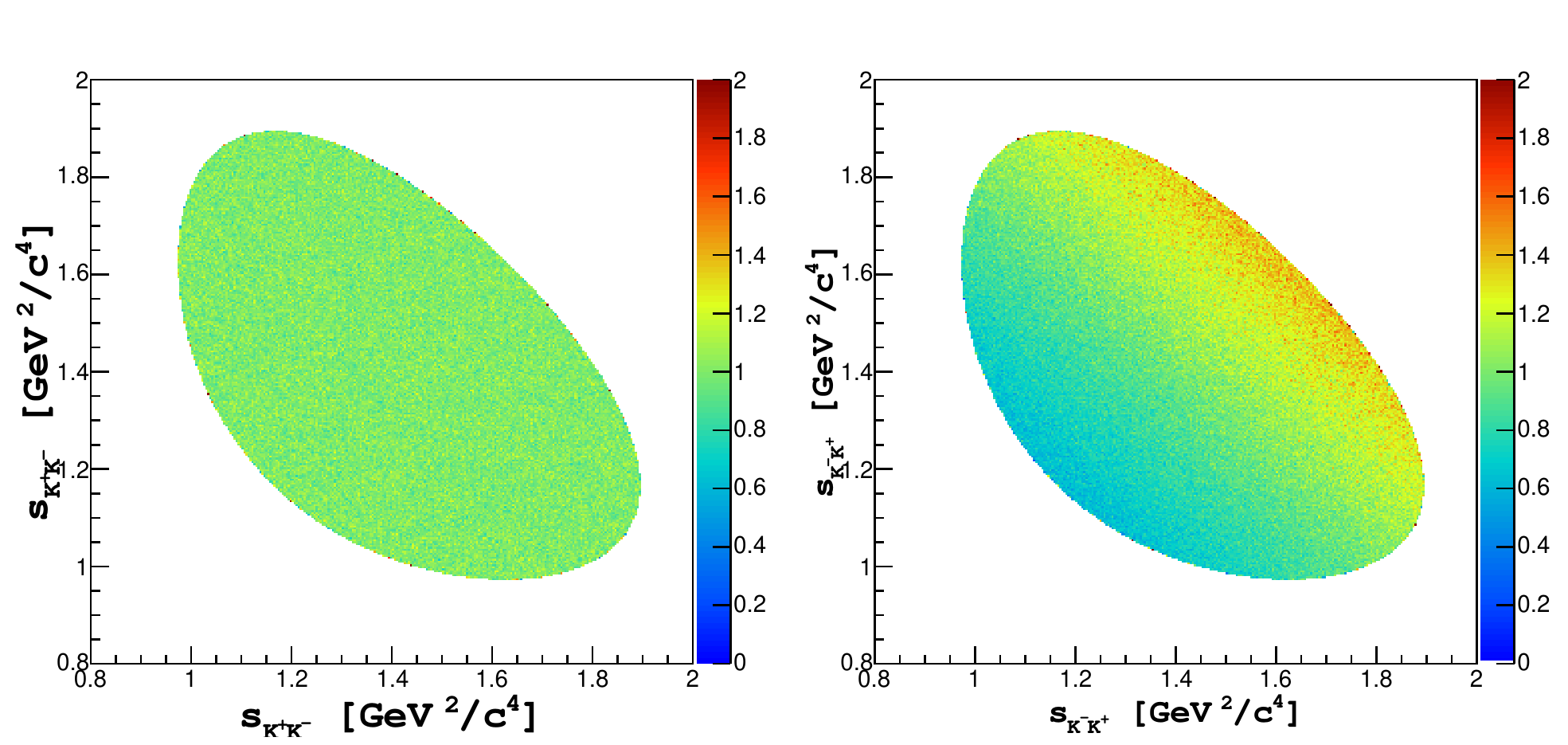}
\caption{Dalitz plot with the MC simulation of: an uniform  nonresonant amplitude  (left);
 the nonresonant component of the Triple-M (right).}
\label{tnr}
\end{figure}

In Figs. \ref{tphi} - \ref{tnr}, we present the results of a simulation to the $D^+\to K^-K^+K^+$ Dalitz
plot distribution for the individual
components of the Triple-M amplitude.
We compare our Triple-M simulation (plots on the right), with the corresponding amplitudes that are commonly used in Dalitz plot analyses (plots on the left):
 a relativistic Breit-Wigner,
for the case of the $\phi$, and  the Flatt\'e function, for the case
of the $f_0$. 

In the case of  $T_{\phi}$ (Fig. \ref{tphi}), the Triple-M and the relativistic BW yield almost identical distributions.
 For the $f_0$, one found  that the singlet and octet
hypothesis yields nearly identical distributions, so only the
singlet case is considered. 
A comparison between $T_{f_0}$ and the Flatt\'e  function, made in Figs.  \ref{tf0}, show that
the two amplitudes result in a somewhat different distributions in the Dalitz plot. The $T_{f_0}$
component yields a distribution which is more concentrated towards the threshold than
that of the Flatt\'e formula. 
The differences between  $T_{NR}$ and the constant nonresonant amplitude, in Fig.  \ref{tnr}, are much larger.
 
Finally, in Fig. \ref{tMMM}, we present the
Dalitz plot distribution with the full Triple-M amplitude as proposed in section V.
One interesting feature is the distribution of events in the $\phi$ region. One of
the lobes is depleted with respect to the other, resulting in a peak and a dip. This
is caused by the interference between the $\phi$ and the $f_0$ components of
the Triple-M amplitude. In this region the strong phases are rapidly changing. 
The resulting distribution of events is very sensitive to the details of the parametrization of these two components.

\begin{figure}[ht] 
\includegraphics[width=0.8\textwidth]{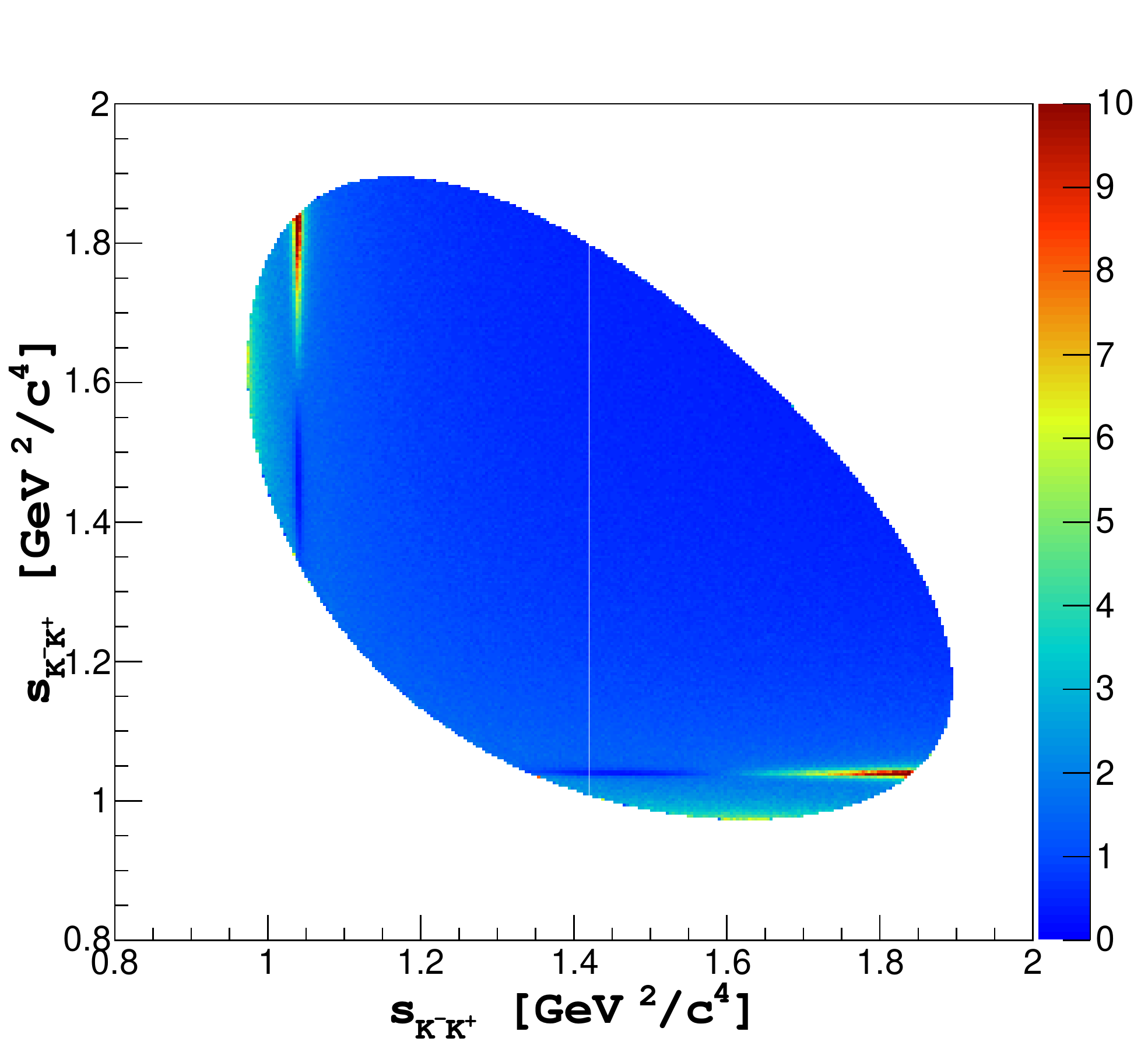}
\caption{Dalitz plot with MC simulation of the full Triple-M amplitude. The
is assumed to be a singlet state. The hypothesis of the $f_0$ 
being an octet state yields a nearly identical distribution.}
\label{tMMM}
\end{figure}

\section{Conclusions}

In this paper, a new approach to decay amplitudes for non-leptonic three-body decays is presented,
applied to the particular case $D^+ \rar K^- K^+ K^+$.
Results, however,  can be easily extended to other decays of charged heavy mesons into
final states containing three charged kaons.
This amplitude relies on matrix elements of weak currents involving multi-meson topologies
and is called multi-meson-model, Triple-M, for short.
The topologies considered are a direct consequence of chiral perturbation 
theory \cite{Wchi, GL, EGPR} and relevant to decays of both
leptons and heavy mesons.
These structures generalize the notion of form factor and, 
at the same time, allow one to go beyond the isobar model, 
often employed in analyses of heavy-meson decays.

We assume that the decay $D^+ \rar K^- K^+ K^+$ is dominated by the direct annihilation
$D^+\rar W^+$ which, subsequently gives rise to the processes shown in Fig. 3. 
The corresponding amplitude is proportional to the product of matrix elements
$\la (KKK)^+ | A^\m|0\ra \la 0|A_\m |M^+\ra$, where $A^\m$ is the axial current.
The Triple-M is composed by a non-resonant term and two resonant contributions,
associated with the $\f$ and the $f_0$.  
The non-resonant amplitude is a direct prediction from chiral symmetry and represented 
by a polynomial, with no free parameters.
It describes a proper three-body interaction, rather than the
of 2+1 decomposition  (two-body subsystem+spectator).
As this contribution involves no loops, it is real for theoretical reasons and,
therefore, adequate for fixing the overall phase of the Triple-M amplitude.

The resonant contributions involve expressions which are very different 
from the $A_k$ used in the isobar model amplitude $A=\sum c_k\,A_k$,
but these expressions yield a similar line shape. 
However, in the Triple-M, the free coefficients $c_k$ are absent,
because the intensity of each resonance is predicted by the underlying dynamics.
In particular, the $\f$ contribution is completely fixed, for 
its intensity is related directly with the decay width into $\Kb K$.
The case of the $f_0$ is different, just because one does not have  precise
values for its mass and couplings.
Therefore,  the three parameters in the amplitude,
namely $m_{f_0}, \,c_d$, and $c_m$,
are left to be determined by fits to data. In the $K^-K^+K^+$ final state one can access
only the tail of the $f_0$, and therefore this channel may not be the best one for
the determination of these three parameters. The decay $D^+_s \to \pi^-\pi^+\pi^+$,
where the $f_0(980)$ is the dominant component, would be the most adequate for
this measurement. It is worth mentioning a recent work \cite{oset} on this
subject, where the $f_0(980)$ line shape is obtained in the context of the
Chiral Unitary theory, from a study of $D^+_s$ decays into $\pi^-\pi^+\pi^+$
and $K^-K^+K^+$.

Our study also encompasses other dynamical effects, representing corrections to
the intermediate $\Kb K$ scattering amplitude,
which were discussed in section IV and found
to be small.
We have left them out of the Triple-M, for the time being, since the ability
of the leading contributions to reproduce data must be tested first.
This kind of testing  would provide important indications about the importance
of effects which are not included in the the present version of the Triple-M,
such as isospin 1 resonances, as well as  dynamical effects  associated with 
processes other than the annihilation diagram.

\section*{ACKNOWLEDGMENTS}
This work  was supported by Conselho Nacional de Desenvolvimento Cient\'ifico e Tecnol\'ogico (CNPq) and M.R.R. would like to thank Funda\c{c}\~{a}o de 
Amparo \`{a} Pesquisa do Estado de S\~{a}o Paulo (FAPESP).

\appendix 

\section{two-meson propagator}
\label{propagator}

The results presented here are conventional and displayed for the sake of completeness. 
One deals with both $S$- and $P$-waves and the corresponding two-meson propagators are associated 
with the integrals
\bea
&&
I_{aa}
= \int \frac{d^4  \ell}{(2\p)^4}
\frac{1} {D_- \; D_+ },
\label{a.1}\\[2mm]
&&
I_{aa}^{\m \n} 
= \int \frac{d^4  \ell}{(2\p)^4}
\frac{\ell^\m  \ell^\n} {D_- \; D_+ },
\label{a.2}\\[2mm]
&& D_\pm = (\ell \sm q/2)^2 \pm  M_a^2 \;,
\nn
\eea
with $q^2=s$.
The integral $I_{aa}$  can be evaluated using dimensional techniques \cite{GL}
and reads \cite{Sch}
\bea
&& I_{aa} = -\, \frac{i}{16 \p^2} \, \lb R + \ln \frac{M_a^2}{\m^2} + 1 \rb - i\, \Ob_{aa}^S \;,
\label{a.3}
\eea
where $R$ is a function of the number of dimensions $n$, which diverges in the limit $n\rar 4$\cite{Sch},
$\m$ is the renormalization scale, and $\Ob_{aa}^S$  the regular part
which, for $s \geq 4M_a^2$, has the form
\bea
&& \Ob_{aa}^S= -\,\frac{1}{16 \p^2} 
\lc 2 \sm \sqrt{\frac{s \sm 4M_a^2}{s}} 
\ln\lb \frac{s- 2M_a^2 + \sqrt{s (s \sm 4M_a^2)}}{2M_a^2}\rb
+ i\, \p\, \sqrt{\frac{s \sm 4 M_a^2}{s}} \;\rc .
\label{a.4}
\eea
In the renormalization process, the divergent factor $R$ is 
replaced by an undetermined constant.
However, there is no need to face this problem here, since we are  concerned just
with on-shell contributions to the propagator, associated with 
its imaginary part.  
Thus, 
\bea
&& \Ob_{aa}^S \rar  -i\,\frac{1}{16 \p}  \, \frac{\sqrt{s \sm 4 M_a^2}}{\sqrt{s}} 
\label{a.5}
\eea

The integral $I_{aa}^{\m\n}$ is evaluated by noting that its Lorentz structure yields
\bea
&&
I_{aa}^{\m \n}  = g^{\m\n} \, A + q^\m q^\n \, B \;,
\label{a.6}
\eea
where $A$ and $B$ are functions of $s$.
Multiplying both eqs.(\ref{a.2}) and (\ref{a.6}), successively  by $2q_\m$ and by $g_{\m\n}$,   
using $2 q\cdot \ell = D_+-D_-$,  $\ell^2 = - \lp s/4 - M_a^2 \rp + (D_+ + D_-)/2$,
and equating results, one finds the conditions 
\bea
&& A + s B = I_a/2 \;,
\nn\\[2mm]
&& 4 A + s B = - (s/4 - M_a^2) \, I_{aa} + I_a \;,
\nn
\eea
with
\bea
&& I_a = \int \frac{d^4  \ell}{(2\p)^4} \frac{1}{\ell^2 - M_a^2}
 = - i\, \frac{M_a^2}{16 \p^2} \, \lb R + \ln \frac{M_a^2}{\m^2} \rb \;,
 \label{a.7}
\eea
which yield 
\bea
&&
I_{aa}^{\m \n}  = \lb g^{\m\n} - \frac{q^\m q^\n}{q^2} \rb 
\lb - \frac{1}{4} \lp q^2 - 4 M_a^2 \rp \;\frac{ I_{aa}}{3} + \frac{I_a}{6} \rb 
+ \frac{q^\m q^\n}{q^2} \, \frac{I_a}{2} \;.
\label{a.8}
\eea
Keeping just the imaginary part, one has 
\bea
&& I_{aa}^{\m\n} \rar \frac{i}{4}\, 
\lb g^{\m\n} - \frac{q^\m q^\n}{s} \rb\; \Ob_{aa}^P \;,
\nn\\[2mm]
&& \Ob_{aa}^P \rar -  \frac{i}{48 \pi} \, \frac{[s \sm 4\,M_a^2]^{3/2}}{\sqrt{s}}  \;.
\label{a.9}
\eea

\section{partially dressed $\f$ propagator}
\label{dp}

The bare $\phi$ propagator, $G_{\a\b\g\d}$, is given by  eq.(A.10) of Ref. \cite{EGPR}.
It  is dressed by $\p \r$ and $\Kb K$ intermediate states and 
the corresponding self-energies are denoted respectively by $\S_{\p\r}$ and $\S_{\Kb K}$.
In this appendix we consider just contributions of the first kind.
The full propagator is given by 
\bea
i\,\D_{\a\b\g\d} &\!=\!&  i\, \D_{\a\b\g\d}^{(0)} + i\, \D_{\a\b\g\d}^{(1)}  
+ i\, \D_{\a\b\g\d}^{(2)}  + i\, \D_{\a\b\g\d}^{(3)} + \cdots  
\label{dp.1}\\
i\,\D_{\a\b\g\d}^{(0)} &\!=\!&  G_{\a\b\g\d} 
\label{dp.2}\\
i\,\D_{\a\b\g\d}^{(1)} &\!=\!&   G_{\a\b ab}\; \lb -i\, \Sigma^{abcd} \rb\; G_{cd\g\d}
\label{dp.3}\\
i\,\D_{\a\b\g\d}^{(2)} &\!=\!&   G_{\a\b ab}\; \lb -i\, \Sigma^{abef} \rb\; G_{efgh}
\; \lb -i\, \Sigma^{gh cd} \rb\; G_{cd\g\d}
\label{dp.4}
\eea

The $\f \p \r$ interaction is extracted from the Lagrangian 
\bea
\cL^{\o_1} &\!=\!& i\,g_1  \; \e^{\m\n\rho\s}\;
\dr^\l \o_{1\, \l\m} \; 
\lb \dr_\n \p^-  \rho_{\rho\s}^+ 
+ \dr_\n \p^+ \rho_{\rho\s}^-
+ \dr_\n \p^0 \rho_{\rho\s}^0 \rb
\label{dp.5}
\eea
where $\o_1= \cos\theta\, \f - \sin\theta \,\o$ is the $SU(3)$ singlet component.
In the sequence, we write $g_\e = g_1\,\cos\theta$.

\begin{figure}[ht] 
\hspace*{-20mm}
\includegraphics[width=6cm,angle=0]{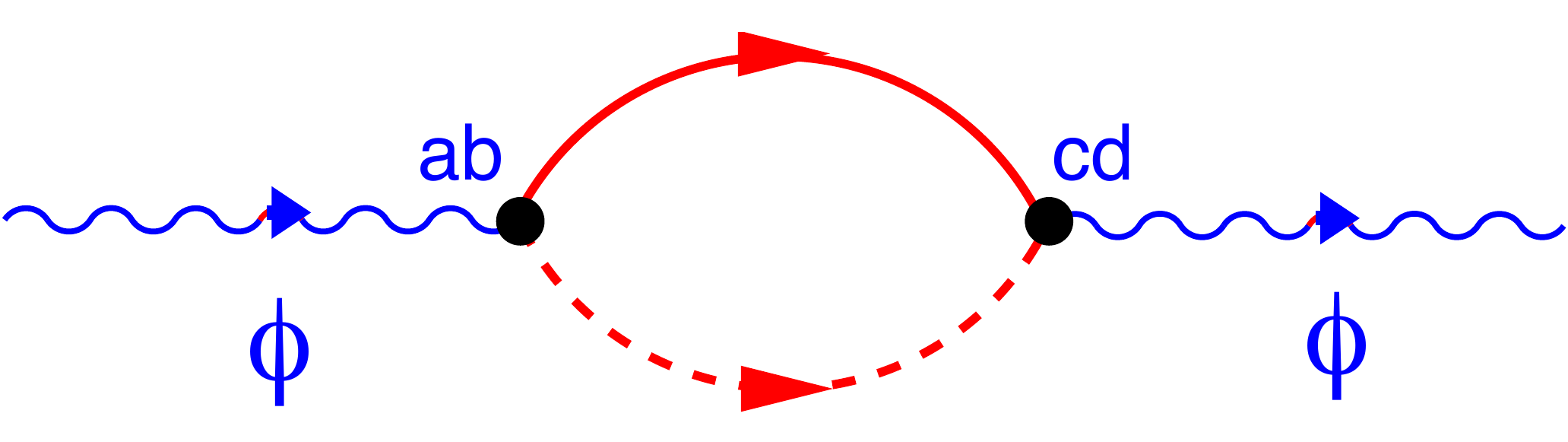}
\caption{Intermediate $\p\r$ contribution to the $\f$ self-energy.}
\label{F1}
\end{figure}

The self energy is given by
\bea
-i\,\Sigma_{\rho\p}^{abcd} &\!=\!& 
\frac{(k^a\,g^{b\m} - g^{a\m}\,k^b)}{2} \; \lb H_{\m\l} \rb \;
\frac{(k^c\,g^{d\l} - g^{c\l}\,k^d)}{2} \;,
\label{dp.6}\\
H_{\m\l}  &\!=\!& 
\lb - 3\, g_\e^2 \; I_{\m\l} \rb \;,
\label{dp.7}\\
I_{\m\l} &\!=\!& \frac{1}{i} \; 
\int \!\! \frac{d^4\ell}{(2\p)^4}\; \frac{p^\m \, p^\l} {p^2 - M_\p^2} \;
\e_{\m\n \x\y} \,G_{\x\y\w\z}(q)\, \e_{\l\xi \w\z} \;,
\label{dp.8}
\eea
with $ p = k/2-\ell, \;\; q =  k/2+\ell$  and $k^2=s\,$.
Using the explicit form of $G_{\x\y\w\z}$ and the definitions
$ D_\p =  p^2 - M_\p^2 \,, \;\; D_\rho = q^2 - \mr2 $, we find
\bea
&& I_{\m\l}  \rar 
\frac{4}{\mr2 } \; 
\int \!\! \frac{d^4\ell}{(2\p)^4}\; \frac{1} {D_\p} \;\frac{1}{D_\r}\;
\label{dp.9}\\
&& \times 
\lc g_{\m\l} \lb -\mr2 \lp M_\p^2 + D_\p  \rp 
+ \frac{1}{4} \,\lp s - M_\p^2 - \mr2 - D_\p - D_\r \rp^2 \rb 
+ \ell_\m \ell_\l \lb k^2 - D_\r \rb \rc \;,
\nn
\eea
where we have used the fact that terms proportional to $k_\m$ and $k_\l$ do not 
contribute to eq.(\ref{dp.6}). 
This integral is highly divergent, but the part regarding the $\p \r$ cut is not.
Terms containing factors $D_\p$ and $D_\r$ in the numerator do not contribute 
to the cut function and the relevant integral is 
\bea
I_{\m\l} \!
 \rar \frac{1}{\mr2 } \; 
\int \!\! \frac{d^4\ell}{(2\p)^4}\; \frac{1} {D_\p \; D_\rho }
\lc  \lb  s^2 - 2\, s \, \lp M_\p^2 + \mr2 \rp + \lp M_\p^2 - \mr2 \rp^2 \rb  g_{\m\l}
+ 4\, s \, \ell_\m \ell_\l   \rc \;.
\label{dp.10}
\eea
Using the definition 
\bea
&& I_{\p\rho} = \int \!\! \frac{d^4\ell}{(2\p)^4}\; \frac{1} {D_\p  \; D_\rho } 
\label{dp.11}
\eea
and the result
\bea
\int \!\! \frac{d^4\ell}{(2\p)^4}\; \frac{\ell_\m \,\ell_\l} {D_\p \; D_\rho }\;
&\!=\! & - \lc \frac{1}{12\,k^2} \;
\lb  s^2 - 2\,  s \, (M_\p^2 + \mr2 ) + (M_\p^2 - \mr2 ) ^2 \rb \, I_{\p\rho} \rc g_{\m\l} 
\nn\\
&\! + \!& \mathrm{term} \, \mathrm{proportional} \, \mathrm{to} \, k_\m\,k_\l  \;,
\label{dp.12}
\eea
the relevant component of $I_{\m\l}$ becomes
\bea
I_{\m\l}  &\! \rar \!&  \lc \frac{2}{3\, \mr2 } \; 
\lb  s^2 - 2\, s \, (M_\p^2 + \mr2 ) + (M_\p^2 - \mr2 ) ^2 \rb \; I_{\p\rho} \rc \; g_{\m\l} \;.
\label{dp.13}
\eea
The on-shell contribution to eq.(\ref{dp.11}) is given by 
\bea
I_{\p\rho} 
&\!=\!&  -\; \frac{1}{16\,\p} \frac{\sqrt{\l_{\p \rho}}}{s} \;,
\label{dp.14}
\eea
with 
$ \l_{\p\rho} = \lb  s^2 - 2\, s \, (M_\p^2 + \mr2 ) + (M_\p^2 - \mr2 ) ^2 \rb 
= 4\, s\, Q_{\p\rho}^2  $, where $Q_{\p\rho}$ is the CM three-momentum.
We then have 
\bea
&& H_{\m\l}  =   g_{\m\l} \, \frac{m_\f}{s} \;\G_\f^{\p\r}(s)\;,
\label{dp.15}\\[2mm]
&& m_\f \, \G_\f^{\p\r} (s) = \frac{g_\e^2}{\p\, m_\r^2} \; s^{3/2} \; Q_{\p\rho}^3 \;.
\label{dp.16}
\eea
Using this result into eq.(\ref{dp.1}) and ressumming  the series, we get
the partially dressed propagator
\bea
i\,\D_{\a\b\g\d}^{\p\r} &\!=\!&  G_{\a\b\g\d} 
\label{dp.17}\\
&\!+\!&
\lb \frac{i \, m_\f\,\G_\f^{\p\r}(s)/s}{ D_\f^{\p\r}(s) } \rb 
\frac{1}{2}
\lb g_\a^d\,k_\b \,k^c + g_\b^c\,k_\a\,k^d - g_\a^c \, k_\b\,k^d - g_\b^d \,k_\a\,k^c \rb \; 
 G_{cd\g\d}  \;,
\nn
\eea
where the denominator $ D_\f^{\p\r}(s) $ is given by 
\bea
D_\f^{\p\r} &\!=\!&  s-m_\f^2 + i\, m_\f\,\G_\f^{\p\r}(s) \;.
\label{dp.18}
\eea
In the evaluation of amplitudes involving a $\Kb(p_1)\,K(p_2)$ vertex, 
one encounters the product 
\bea 
i\, \D_{\a\b\g\d} \, \lp p_1^\g p_2^\d - p_2^\g p_1^\d \rp   &\!=\!&  
-\, \frac{2\,i}{D_\f^{\p\r}(s)}\; 
\lb p_{1\a} p_{2\b} - p_{2\a} p_{1\b} \rb \;.
\label{dp.19}
\eea

\section{$\bar{K}K$ amplitude}
\label{kk}

We construct the $\bar{K}K$ amplitude by deriving interaction kernels $\cK$ from chiral Lagrangians  \cite{EGPR}
involving resonances and iterating them in the $s$-channel, by means of two-kaon loops.  
In the treatment of the $S$-channel, one considers the coupling of $\p\p$ and $\Kb K$ states,
whereas in the $P$-channel we distinguish $K^- K^+$ and $\Kb^0 K^0$ intermediate states.

\subsection{coupled channel formalism}

Basic formulae given here apply to both $S$- and $P$- channels. 
In the absence of coupling, the amplitudes $t$ are given as sums of Dyson series 
\bea
t_{11}&\!=\!& \frac{\cK_{11}}{1 + \Omega_{11}\,\cK_{11}}  \;,
\label{c.1}\\[4mm]
t_{22}&\!=\!& \frac{\cK_{22}}{1 + \Omega_{22}\,\cK_{22}}  \;
\label{c.2}
\eea
where the functions $\Omega_{ii}$ describe the propagation of two intermediate
meson of mass $M_i$.

In the diagonal channels, one can construct effective kernels $\cKb$, given by
\bea
\cKb_{11} &\!=\!& \cK_{11}-\cK_{12}\, 
\lb 1 - \Omega_{22}\,t_{22} \rb \, \Omega_{22} \, \cK_{21} \;,
\label{c.3}\\[4mm] 
\cKb_{22} &\!=\!& \cK_{22}-\cK_{21}\, 
\lb 1 - \Omega_{11}\,t_{11} \rb \, \Omega_{11} \, \cK_{12} \;,
\label{c.4}
\eea

The diagonal coupled amplitudes $T$  are obtained by iterating the effective kernels
and read
\bea
T_{11}&\!=\!& \frac{\cKb_{11}}{1 + \Omega_{11}\,\cKb_{11}}  \;,
\label{c.5}\\[4mm]
T_{22}&\!=\!& \frac{\cKb_{22}}{1 + \Omega_{22}\,\cKb_{22}}  \;,
\label{c.6}
\eea
and their explicit forms are
\bea
T_{11}&\!=\!& \frac{\cK_{11}+\Omega_{22}\,||\cK||}
{1 + \Omega_{11}\,\cK_{11} + \Omega_{22}\,\cK_{22} 
+ \Omega_{11}\, \Omega_{22}\, ||\cK||}  \;,
\label{c.7}\\[4mm]
T_{22}&\!=\!& \frac{\cK_{22} + \Omega_{11}\,||\cK||}
{1 + \Omega_{11}\,\cK_{11} + \Omega_{22}\,\cK_{22} 
+ \Omega_{11}\, \Omega_{22}\, ||\cK||}  \;,
\label{c.8}
\eea
with
\beq
||\cK||= \cK_{11}\,\cK_{22}- \cK_{12}^2 \;.
\label{c.9}
\eeq
The off-diagonal term is
\bea
T_{12}&\!=\!& \frac{\cK_{12}}
{1 + \Omega_{11}\,\cK_{11} + \Omega_{22}\,\cK_{22} 
+ \Omega_{11}\, \Omega_{22}\, ||\cK||}  \;.
\label{c.10}
\eea

\subsection{$S$-channel amplitude}

In the treatment of the $S$-channel, one needs three basic kernels, namely
\\
$\cK_{11} \rar [\p\p \lrar \p\p]$,
$\cK_{12} \rar [\p\p \lrar KK]$,
$\cK_{22} \rar [KK \lrar KK]$. 

As the role of the $f_0$ in the $SU(3)$ structure is not clear, we allow it to be either a singlet,
with mass $m_0$, or a member of an octet, with mass $m_8$.

The relevant kernels read
\bea
\cK_{11}  &\!=\! & \frac{1}{F_\p^2} \lb\frac{2s - M^2_\p}{3} \; 
- \frac{4}{3}\, \frac{G_\p^2}{s-m_0^2} 
- \frac{2}{3}\;\frac{G^2_\p}{s - m_8^2}\;,\rb 
\label{c.11} \\[2mm] 
\cK_{22}  &\!=\!&  \frac{1}{F_K^2} \lb \frac{3\,s}{8} 
- \frac{4}{3}\, \frac{G_K^2}{s-m_0^2} 
- \frac{1}{6}\;\frac{G^2_K}{s - m_8^2} \rb\;,
\label{c.12} \\[2mm]
\cK_{12}&\!=\! & \frac{1}{F_\p F_K} \lb \frac{s}{4} 
- \frac{4}{3}\, \frac{G_\p\,G_K}{s-m_0^2}  
+ \frac{1}{3}\;\frac{G_\p\,G_K}{s - m_8^2} \rb\;,  
\label{c.13}
\eea
with
\bea
G_\p &\!=\! & \frac{1}{F_\p}\lb c_d\,s - 2\,(c_d \sm c_m)\,M_\p^2 \rb   \;,
\label{c.14}\\[2mm]
G_K &\!=\! & \frac{1}{F_K} \lb c_d\,s - 2\,(c_d \sm c_m)\,M_K^2 \rb   \;,
\label{c.15}
\eea 
where $c_d$ and $c_m$ are coupling constants \cite{EGPR} and we have used 
$\tilde{c}_i = c_i/\rth$.
 
In the K-matrix approximation, one needs just the  on-shell component of the two-meson 
propagators, which are given by 
\bea
\Omega_{\p\p} &\!=\!& - \lb \frac{3}{2}\rb \,\frac{i}{16\,\p} \;\frac{\sqrt{s-4\,M_{\p}^2}}{\sqrt{s}}  
= -\,\frac{3 i}{16\,\p} \;\frac{Q_{\p\p}}{\sqrt{s}}  \;,
\label{c.16}\\[2mm]
\Omega_{KK} &\!=\!& -\lb \frac{4}{2}\rb \,\frac{i}{16\,\p} \;\frac{\sqrt{s-4\,M_{K}^2}}{\sqrt{s}}
= -\,\frac{i}{4\,\p} \;\frac{Q_{KK}}{\sqrt{s}}  \;,
\label{c.17}
\eea
which include both the multiplicities of intermediate states and the symmetry factor $1/2$.

The scattering amplitudes can be obtained by using eqs.(\ref{c.11}-\ref{c.17}) into 
results (\ref{c.7}-\ref{c.10}).
However, this yields expressions which are rather cumbersome.
In order to simplify the results, we neglect contact interactions in eqs.( \ref{c.11}-\ref{c.13}) 
and assume the $f_0(980)$ to be either a $SU(3)$ singlet or a member of an octed.
These choices are indicated, respectively, by labels $0$ and $8$.

In the singlet case, one has 
\bea
&& T_{11}^0 = - \frac{4\,G_\p^2}{3 F_\p^2} \; \frac{1}{D_0} \;,
\hspace*{10mm}
 T_{22}^0 = - \frac{4\,G_K^2}{3 F_K^2} \; \frac{1}{D_0} \;,
\hspace*{10mm}
T_{12}^0 = - \frac{4\,G_\p \, G_K}{3 F_\p F_K} \; \frac{1}{D_0} \;,
\label{c.18}\\[2mm]
&& D_0 = \lp s-m_{f_0}^2 \rp + i\, m_{f_0} \, \G_0 \;,
\label{c.19}\\[2mm]
&& m_{f_0}\,\G_0 = \frac{G_\p^2}{4\p F_\p^2} \, \frac{Q_{\p\p}}{\sqrt{s}}
+ \frac{G_K^2}{3\p F_K^2} \, \frac{Q_{KK}}{\sqrt{s}} \;.
\label{c.20}
\eea
Alternatively, for the octet, one finds
\bea
&& T_{11}^8 = - \frac{2\,G_\p^2}{3 F_\p^2} \; \frac{1}{D_8} \;,
\hspace*{10mm}
 T_{22}^8 = - \frac{G_K^2}{6 F_K^2} \; \frac{1}{D_8} \;,
\hspace*{10mm}
T_{12}^8 =  \frac{G_\p \, G_K}{3 F_\p F_K} \; \frac{1}{D_8} \;,
\label{c.21}\\[2mm]
&& D_8 = \lp s-m_{f_0}^2 \rp + i\, m_{f_0} \, \G_8 \;,
\label{c.22}\\[2mm]
&& m_{f_0}\,\G_8 = \frac{G_\p^2}{8\p F_\p^2} \, \frac{Q_{\p\p}}{\sqrt{s}}
+ \frac{G_K^2}{24\p F_K^2} \, \frac{Q_{KK}}{\sqrt{s}} \;.
\label{c.23}
\eea
%

\subsection{$P$-channel amplitude}

In the $P$-channel, we consider the kernels
\\
$\cK_{11} \rar [K^- K^+\lrar K^- K^+]$,
$\cK_{12} \rar [\Kb^0 K^0\lrar \Kb^0 K^0]$,
$\cK_{22} \rar [K^- K^+\lrar \Kb^0 K^0]$.

They are related to  tree amplitude is given by $(t-u)\,\cK$ 
and given by 
\bea
&& \cK_{11}  = \cK_{22}  = - \cK_{12} =  \cK_0^P/2 \;,
\label{c.24}
\eea 
with
\bea
\cK_0^P(s) &\! = \!&  \frac{3}{4\,F_K^2}
- \lb \sin^2 \theta\; \frac{3\,G_V^2}{2\,F_K^4}\rb \;
\frac{s}{D_\f^{\p\r}(s)} \;.
\label{c.25}
\eea
In the evaluation of amplitudes, it is convenient to express it as 
\bea
\cK_0^P(s)  &\!=\!& \frac{N_0^P(s)}{D_\f^{\p\r}(s)} \;,
\label{c.26}\\[4mm]
N_0^P(s) &\!=\!& - \lc \sin^2\th \;\frac{3\,G_V^2}{2\,F_K^4} \,s
- \,\frac{3\, D_\f^{\p \r}}{4\,F_K^2}  \rc  \;,
\label{c.27}
\eea
The two-kaon propagators are  
\bea
\Omega_{11}^P &\!=\!& -\,\frac{i}{48\,\p} \;\frac{[s-4\,M_{K^+}^2]^{3/2}}{\sqrt{s}}  
= -\,\frac{i}{6\,\p} \;\frac{|Q_{11}|^3}{\sqrt{s}}  \;,
\label{c.28}\\[4mm]
\Omega_{22}^P &\!=\!& -\,\frac{i}{48\,\p} \;\frac{[s-4\,M_{K^0}^2]^{3/2}}{\sqrt{s}}
= -\,\frac{i}{6\,\p} \;\frac{|Q_{22}|^3}{\sqrt{s}}  \;.
\label{c.29}
\eea

These results yield the various components of the $KK$ amplitude, which are given by
\bea
T_{11} = T_{22} = - T_{12} = T_0^P/2  \; .
\label{c.30}
\eea
Its explicit form is 
\bea
T_0^P(s) &\!=\! & \frac{\cK_0^P}{1 + [\Omega_{11}^P +  \Omega_{22}^P]  \,\cK_0^P} 
= \frac{N_0^P(s)}{D_\f(s)} 
\label{c.31}\\[2mm]
D_\f(s) &\!=\!&  s - m_\f^2  + i\, m_\f \, \G_\f(s) 
- i\, \frac{1}{6\p} \,\frac{3}{4 F_K^2} \, \frac{D_\f^{\p\r}(s)}{\sqrt{s}}\; \lp  Q_{11}^3+ Q_{22}^3\rp 
\nn\\[4mm]
m_\f \, \G_\f(s) &\!= \!& 
\lb  \frac{1}{6\p}\, \sin^2\th \, \frac{3\, G_V^2}{2\, F_K^4}\; \sqrt{s}\; \lp Q_{11}^3 + Q_{22}^3\rp  
+ \frac{1}{\p}\,\frac{g_\e^2}{m_\rho^2} \; s^{3/2}\; Q_{\p \rho}^3 \rb 
\label{c.32}
\eea
For $s=m_\f^2$, the imaginary part of $D_\f$ becomes equal to $m_\f\;\Gamma_\f$.
Using $\Gamma_\f = \Gamma_{\p\rho} +\Gamma_{KK}$, we determine the coupling 
constants, which are given by
\bea
&& \frac{\sin^2\th}{6\,\p} \, \frac{3\, G_V^2}{2\, F_K^4} = \frac{\Gamma_{KK}}{( \Qt_{11}^3+\Qt_{22}^3)  } \;,
\label{c.33}\\[2mm]
&& \frac{1}{\p}\,\frac{g_\e^2}{m_\rho^2} = \frac{\Gamma_{\p\rho}}{m_\f^2 \, \Qt_{\p\rho}^3} \;,
\label{c.34}
\eea 
where $\Qt \equiv  Q( s=m_\f^2)$.
These results allow eqs.(\ref{c.27}) and (\ref{c.32})  to be written as 
\bea
&& N_0^P = - \frac{6\p\,\Gamma_{KK}}{(\Qt_{cc}^3+\Qt_{nn}^3)} \,s
+ \,\frac{3\,D_\f^{\p\r}}{4\,F_K^2}  \;,
\label{c.35}\\[4mm]
&& D_\f^{\p\r} = s - m_\f^2  + i \,\G_{\p\r} \; \frac{s^{3/2}}{m_\f^2} \; \frac{Q_{\p\r}^3}{\Qt_{\p\r}^3} \;,
\label{c.36}\\[4mm]
&& m_\f \, \G_\f (s) = 
 \lb \Gamma_{KK}\, \sqrt{s}\; \frac{( Q_{11}^3 + Q_{22}^3)} {( \Qt_{11}^3 + \Qt_{22}^3)}  
+\Gamma_{\p\rho} \;\frac{s^{3/2}}{m_\f^2} \; \frac{Q_{\p \rho}^3}{\Qt_{\p \rho}^3} \rb \;.
\label{c.37}
\eea

\section{production amplitudes}
\label{pa}

\subsection{$S$-channel}
\begin{figure}[ht] 
\includegraphics[width=2cm,angle=90]{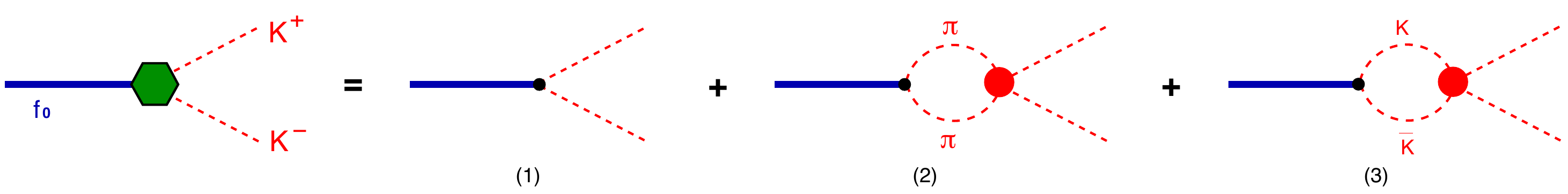}
\caption{$f_0$ production  with $K\bar{K}$ and $\p\p$ couple channel contribution.}
\label{FigS}
\end{figure}

The production amplitude for a final charged $K^-K^+$ pair in the S-channel 
is defined in Fig. \ref{FigS} and includes $\p\p$ and $\Kb K$ loops coupled to the $f_0(980)$. 

In the case of a singlet $f_0$, the diagrams above are described by
\bea
&& i\,\Pi_{(1)}^0 +  i\,\Pi_{(2)}^0 +  i\,\Pi_{(3)}^0 
=  -\, \frac{2\,G_K}{\rth \,F_K}\,\frac{1}{s- m_0^2}
\lb 1 + \frac{\Omega_{11}}{D_0} \, \frac{4\;G_\p^2}{3\,F_\p^2}\,
+ \frac{\Omega_{22}}{D_0} \, \frac{4\;G_K^2}{3\,F_K^2}\, \rb 
\nn\\[2mm]
 && = i\,\Pi^0 = - \, \frac{2\,G_K}{\rth \, F_K}\,\frac{1}{D_0} \;,
\label{d.1}
\eea
with $D_0$ given by eq.(\ref{c.19}).

If the $f_0$ is as a member of an octet, we get
\bea
&& i\,\Pi_{(1)}^8 +  i\,\Pi_{(2)}^8 +  i\,\Pi_{(3)}^8 
=  \frac{G_K}{\sqrt{6} \,F_K}\,\frac{1}{s- m_8^2}
\lb 1 + \frac{\Omega_{11}}{D_8} \, \frac{2\;G_\p^2}{3\,F_\p^2}\,
+ \frac{\Omega_{22}}{D_8} \, \frac{G_K^2}{6\,F_K^2}\, \rb 
\nn\\[2mm]
 && = i\,\Pi^8 =  \frac{G_K}{\sqrt{6} \, F_K}\,\frac{1}{D_8} \;,
\label{d.2}
\eea
with $D_8$ as in eq.(\ref{c.22}).

\subsection{$P$-channel}
\begin{figure}[ht] 
\includegraphics[width=2cm,angle=90]{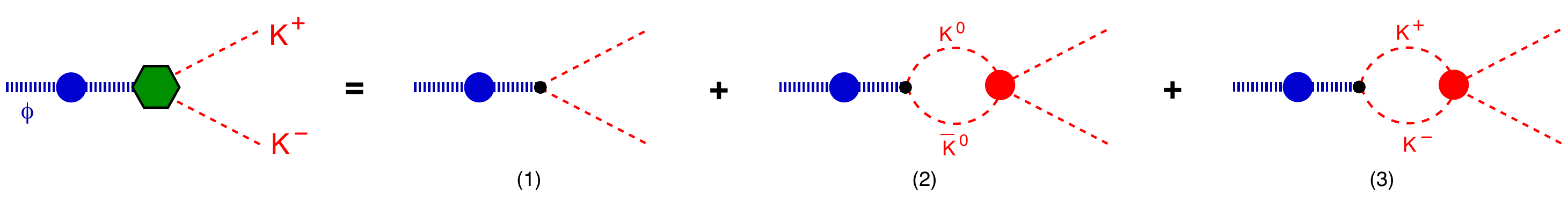}
\caption{$\phi$ production with  propagator dressed by $\p\r$ and,  $K^+K^-$ and $K_0 \bar{K_0}$ couple channel  contribution.}
\label{FigP}
\end{figure}

The production amplitude $\Pi_{\a\b\g\d}$ is defined by the processes 
indicated in Fig. \ref{FigP}.
Explicit evaluation yields
\bea
&& i\,\Pi_{(1)\,\a\b}  +  i\,\Pi_{(2)\,\a\b}  +  i\,\Pi_{(3)\,\a\b}  
\nn\\[2mm]
&& = 2\, i \,\lb \sin\theta\,\frac{\rth \, G_V} {2 F_K^ 2} \rb \;
\frac{1}{D_\f^{\p\r}} 
\lb 1  - \Omega_{11}^P \, T_{11}^P  - \Omega_{22}^P \, T_{22}^P \,\rb 
\lb p_{1\a} \, p_{2\b} - p_{2\a} \, p_{1\b}\rb 
\nn\\[2mm]
&& = i\,\Pi_{\a\b} = 2\, i \,\lb \sin\theta\,\frac{\rth \, G_V} {2 F_K^ 2} \rb \;
\frac{\lb p_{1\a} \, p_{2\b} - p_{2\a} \, p_{1\b}\rb}{D_\f (s)} \;.
\label{d.3}
\eea

\section{individual tree currents}
\label{itc}

Individual contributions from the direct reading of the 
diagrams of Fig. \ref{FA3} to the matrix element
$A^\m \equiv \la K^-(p_1) K^+(p_2) K^+(p_3)|A^\m|\,0\,\ra$
are given below. 
\bea
%
%
%
A_{(1)}^\m &\!=\!& i\, \lb \frac{2\, \rtw}{3 F_K}\rb  \; 
(2\,p_1 \sm p_2 \sm 2\,p_3)^\m \;,
\label{ic.1}\\[2mm]
%
%
%
A_{(2)}^\m &\!=\!& - i\, \lb \frac{2 \rtw}{3 F_K}\rb  \; 
\frac{P^\m}{P^2 \sm M_K^2}\;
\lb  p_1 \cd (p_2 \sp p_3) - 2 \, p_2 \cd p_3 + M_K^2 \rb \;,
\label{A.2}\\[2mm]
%
%
%
A_{(3)}^\m &\!=\!& - i\, \lb \sin^2\theta\, \frac{3 \rtw\, F_V G_V}{2 F_K^3} \rb \; 
\frac{\lb \, P \cd p_2 \,p_1^\m - P \cd p_1 \, p_2^\m \, \rb }
{D_\f^{\p\r }(m_{12}^2)} \;,
\label{A.3}\\[2mm]
%
A_{(4)}^\m &\!=\!& i\, \lb \sin^2\theta \,\frac{3\sqrt{2} \, G_V^2}{F_K^3} \rb \;
\frac{\lb \, p_2\cd p_3 \, p_1^\m - p_1 \cd p_3 \, p_2^\m \, \rb }
{D_\f^{\p\r}(m_{12}^2)} \;,
\label{A.4}\\[2mm]
%
%
%
A_{(5)}^\m &\!=\!& - i\,\lb \sin^2\theta \, \frac{3\sqrt{2}\, G_V^2}{F_K^3} \rb \,
\frac{P^\m}{P^2 \sm M_K^2}\;
\frac{\lb\, P \cd p_1 \, p_2 \cd p_3 - P \cd p_2 \, p_1 \cd p_3\,\rb}
{D_\f^{\p\r}(m_{12}^2)} \;,
\label{A.5}\\[2mm]
%
%
%
A_{(6)}^\m &\!=\! &  i\,\lb \frac{2\,\gamma_n\,\rtw\,c_d}{3\ F_K^3} \rb  \, 
p_3^\m \; \frac{\lb c_d \, p_1 \cd p_2 + c_m \, M_K^2 \rb }{m_{12}^2 - m_{f_0}^2} \;,
\label{A.6}\\[2mm]
%
%
A_{(7)}^\m &\!=\! &  - i\,\lb \frac{2\,\gamma_n\,\rtw }{3\ F_K^3} \rb  \;
\frac{P^\m}{P^2 - M_K^2} \, \lb c_d \, P \cd p_3 - c_m \, M_K^2 \rb \; 
\frac{\lb c_d \, p_1 \cd p_2 + c_m \, M_K^2 \rb }{m_{12}^2 - m_{f_0}^2} \;.
\label{A.7}
\eea


\end{document}